\documentclass[nofootinbib,amsmath,amssymb,showpacs,showkeys,aps,reprint]{revtex4-1}
\usepackage[utf8,latin1]{inputenc}
\usepackage{graphicx}
\usepackage{dcolumn}
\usepackage[dvipsnames]{xcolor}
\usepackage[T1]{fontenc}

\usepackage{mathrsfs}  
\usepackage{cases}
\usepackage{bm}
\usepackage{academicons}
\usepackage{mathtools, nccmath}
\usepackage{fancyhdr}
\usepackage{tikz,xcolor}
\newcommand{\qm}[1]{``#1''}
\usepackage{tensor}
\usepackage[normalem]{ulem}
\usepackage{lipsum}
\usepackage{soul}
\usepackage{cancel}
\usepackage{stackengine,scalerel}
\usepackage{mathabx}
\usepackage{hyperref}
\usepackage{tabularx}
\hypersetup{colorlinks, linkcolor={red},citecolor={blue},urlcolor={blue}}

\definecolor{lime}{HTML}{A6CE39}
\DeclareRobustCommand{\orcidicon}{
	\begin{tikzpicture}
	\draw[lime, fill=lime] (0,0) 
	circle [radius=0.16] 
	node[white] {{\fontfamily{qag}\selectfont \tiny ID}};
	\draw[white, fill=white] (-0.0625,0.095) 
	circle [radius=0.007];
	\end{tikzpicture}
	\hspace{-2mm}
}

\newcommand{\dd}{{\rm d}}
\newcommand{\ii}{{\rm i}}
\newcommand{\ee}{{\rm e}}
\newcommand{\F}{\mathcal{F}}

\fancypagestyle{plain}{%
  \fancyhf{}
  \fancyfoot[C]{\iffloatpage{}{\thepage}}
  }
\foreach \x in {A, ..., Z}{%
	\expandafter\xdef\csname orcid\x\endcsname{\noexpand\href{https://orcid.org/\csname orcidauthor\x\endcsname}{\noexpand\orcidicon}}
}

\begin{document}

\title[Null geodesics, causal structure, and matter accretion \\ in Lorentzian-Euclidean black holes]{Null geodesics, causal structure, and matter accretion \\ in Lorentzian-Euclidean black holes}

\author{Salvatore Capozziello\orcidB{}$^{1,2,3}$}
\email{capozziello@na.infn.it}
\author{Emmanuele Battista\orcidA{}$^{4}$\vspace{0.5cm}}\email{ebattista@lnf.infn.it} \email{emmanuelebattista@gmail.com}
\author{Silvia De Bianchi\orcidC{}$^{5}$} \email{silvia.debianchi@unimi.it}

\affiliation{$^1$ Dipartimento di Fisica ``Ettore Pancini'', Complesso Universitario 
di Monte S. Angelo, Universit\`a degli Studi di Napoli ``Federico II'', Via Cinthia Edificio 6, 80126 Napoli, Italy,\\
$^2$ Istituto Nazionale di Fisica Nucleare, Sezione di Napoli, Complesso Universitario 
di Monte S. Angelo, Via Cinthia Edificio 6, 80126 Napoli, Italy,\\
$^3$ Scuola Superiore Meridionale, Largo San Marcellino 10, 80138 Napoli, Italy,\\
$^4$ Istituto Nazionale di Fisica Nucleare, Laboratori Nazionali di Frascati, 00044 Frascati, Italy, \\
$^5$ Dipartimento di  Filosofia, Universit\`a degli Studi di Milano,  via Festa del Perdono 7, 20122 Milano, Italy. }

\date{\today} 

\begin{abstract}

Recently, we  introduced the Lorentzian-Euclidean black hole, a static and spherically symmetric solution of vacuum Einstein equations that exhibits a change in metric signature across the event horizon. In this framework,  the analysis of radial trajectories of freely falling bodies proves that the central singularity can be avoided via a mechanism we interpret as atemporality,  which  is responsible for the shift of the time variable from real to imaginary values. 

In this paper, we further explore this model by first examining  the behavior of null geodesics. Our investigation requires  a set of signature-adaptive  coordinate changes  that generalize the  local Lorentz transformations underlying General Relativity. We find that photon orbits, like their massive counterparts, cannot traverse the event horizon, thereby strengthening the previous result on the impossibility to reach  the $r=0$ singularity. Additionally, we discuss  the causal structure of the spacetime,  provide the corresponding Penrose diagram, and analyze the  process of matter accretion in the outer region of the black hole.

\end{abstract}

\maketitle

\section{Introduction}

Black holes are  among the most fascinating predictions of Einstein theory. Their existence is indicated  by  observations  of  gravitational waves from  inspiralling compact binaries provided by the LIGO-Virgo-KAGRA network   \cite{LIGOScientific2016,LIGOScientific2025},    the shadows of the supermassive black holes  M87$^*$ and SgA$^*$ captured by the Event Horizon Telescope collaboration \cite{EventHorizonTelescope2019,EventHorizonTelescope2022}, and the electromagnetic radiation emitted by  accretion disks  \cite{Nampalliwar2018}. 

The occurrence of black holes brings to light a  key facet of General Relativity: the presence of singularities. The first exact solution to the Einstein field equations, discovered by Schwarzschild in 1916,  describes the gravitational field in the exterior region of a static, spherically symmetric source and serves as a foundational paradigm for black holes.  The resulting geometry exhibits two notable features: a central curvature singularity, characterized by curvature invariants (such as the Kretschmann scalar) attaining unboundedly large values at $r=0$, and an event horizon at $r=2M$, which establishes the concept of  \qm{no-escape region} for black holes, as it represents a one-way causal boundary in spacetime from which nothing, not even light, can escape.

Initially, spacetime singularities, such as  the one  in the  Schwarzschild solution and those  later identified in the dynamical Oppenheimer-Snyder model or the Kerr geometry \cite{Poisson2009}, were  thought to be artifacts of the high degree of symmetry in the spacetime and were expected to be absent in more realistic scenarios. 
This perspective changed dramatically with the advent of the  Penrose and Hawking singularity theorems \cite{Hawking1973-book}. In particular,  Penrose's 1965 theorem \cite{Penrose1965} was the first one which did not rely on any assumption regarding  symmetries and  proved that singularities are a generic feature of gravitational collapse that must inevitably arise shortly after the emergence of a closed trapped surface \cite{Landsman2021,Landsman2022} (i.e., a spacelike two-surface lying in such a strong gravitational field that even the outgoing light rays are dragged back and forced to converge \cite{Hawking1973-book}).

Strictly speaking, singularity theorems do not demonstrate the existence of black holes, but rather show that, under reasonable hypotheses, the causal geodesics of  spacetime manifolds cannot be extended to arbitrary values of the affine parameter.  To infer  the formation of a black hole from this property, referred  to as geodesic incompleteness, one must  rely on  the so-called  cosmic censorship conjecture \cite{Penrose1969}. First postulated by Penrose, this asserts that singularities stemming from collapsing bodies are not naked but are always hidden within black holes, and thus, not visible from infinity \cite{Wald-book1984}.

Singularities represent a breakdown of General Relativity, pointing out  its inability  to provide boundary conditions  for the field equations at the singular points \cite{Hawking1979-path}. Such regions  signal regimes in which the classical description fails, as they involve infinite physical quantities, lie beyond the observational reach, and undermine the determinism and completeness of the theory. Their very existence suggests that General Relativity, while successful in many domains, may not be the ultimate model of gravity. Such issues, along with  recent and future observations leaving room for alternative structures to ordinary black holes \cite{Berti2015,Barack2018,LISA2022,Herdeiro2022},  have sparked significant interest in  exploring modified and extended gravity theories, which seek for addressing  shortcomings of Einstein theory while preserving its experimentally well-tested predictions  \cite{Capozziello:2011et}. Setting aside horizonless exotic compact objects that could mimic black holes in their strong-gravity phenomenology \cite{Cardoso-Pani2019},  several proposals for regular black holes that lack the central singularity  have been advanced in the literature.  These include some higher-dimensional prototypes naturally arising from string theory (such as   Myers-Perry black holes, which in odd dimensions and with specific spin configurations exhibit no singularities in the domain $r\geq 0$ \cite{Emparan2008,Myers2011}), as well as several four-dimensional models   (see e.g. Refs. \cite{Ansoldi2008,Lan2023,Carballo-Rubio2025} for  reviews). 

The first static, spherically symmetric, regular solution of Einstein equations with an event horizon was achieved by Bardeen  by replacing the mass of the Schwarzschild black hole with an $r$-dependent function  \cite{Bardeen1968}. 
Some years later, building on some very early ideas dating back to the seminal works by Sakharov \cite{Sakharov1966} and  Gliner \cite{Gliner1966},  a nonsingular modified Schwarzschild geometry, known as Dymnikova black hole, was worked out \cite{Dymnikova1992}.  This object is generated by the spherically symmetric vacuum and features a regular, asymptotically de Sitter core as the radial coordinate $r $ goes to zero.  
Along similar lines, other spherically symmetric models have also been devised \cite{Dymnikova2003,Hayward2005} (see also e.g. Refs. \cite{Fathi2021,Molina2021,Macedo2024} for a detailed analysis of the thermodynamic aspects).

Following the outcome of Ref. \cite{Ayon-Beato2000}, which indicated that  regular black holes can be interpreted as  gravitational fields generated by nonlinear electric or magnetic monopoles, numerous papers  appeared in the literature formalizing these entities within Einstein gravity coupled to nonlinear electrodynamics (see e.g. Refs. \cite{Fan2016,Bokulic2022,Canate2022,Guo2023,DeFelice2024}). Furthermore, the central singularity of black holes can  be resolved by taking into account quantum gravitational effects \cite{Bonanno2000,Gambini2008,Platania2019,Babichev2020b,Brahma2020,Roupas2022,Eichhorn2022,Chen2023,Zhang2023,Liu2024b,Lin2024,Xamidov2025}, as well as extended gravity frameworks  (see e.g. Refs. \cite{Bambi2016,Moffat2018,Jusufi2018,Bouhmadi-Lopez2020,Nashed:2021sji,Sau2022,Malekolkalami2023,Capozziello:2023vvr,Gohain2024a,Gohain2024b,Liu2024,Misyura2024,Vertogradov2024,Ovgun2024,Koivisto2024,Vertogradov2025} and references therein), where, among the others, a crucial role is fulfilled by the so-called hairy black holes (see e.g. Refs. \cite{Dias2018,Ovalle2023,Mustafa2024c}).

Recently, inspired by signature-changing models in quantum cosmology that prevent the formation of the Big-Bang singularity  \cite{Gibbons1990,Dereli1993}, we  developed a static, spherically symmetric black hole solution whose signature changes upon crossing the event horizon \cite{Capozziello2024} (see also Ref. \cite{DeBianchi2025}). In the outer region $r>2M$, the metric $g_{\mu \nu}$ assumes the ordinary Lorentzian signature, with the  geometry displaying the same properties as in the standard Schwarzschild one, while at $r=2M$, $g_{\mu \nu}$ becomes degenerate, and for $r<2M$, it admits a Euclidean structure with an ultrahyperbolic signature. We  termed this solution  Lorentzian-Euclidean black hole. 

In this framework, the event horizon constitutes the so-called change surface, i.e., the hypersurface where $g_{\mu \nu}$  undergoes the signature transition. This variation in the metric configuration entails the presence of discontinuities, which give rise to ill-defined Dirac-delta-like singularities in both  the Riemann and Weyl tensors that  can be physically
related to  the presence of a surface layer (or thin shell) and a gravitational
shock wave \cite{Barrabes-book2003,Poisson2009}. We  dealt with these terms by borrowing a technique from gravitational-wave theory and compact binaries dynamics known as the Hadamard \emph{partie finie} regularization scheme \cite{Blanchet2000}. By applying this method, we  regularized both the Riemann and Weyl tensors, which have been shown to contain no distributional piece, and  the  Ricci tensor, which has been found to vanish everywhere in the spacetime. This approach allowed us to conclude that the Lorentzian-Euclidean Schwarzschild metric represents a well-defined signature-changing solution of the vacuum Einstein field equations.
  
One of the most intriguing aspects of the regularized geometry is that both freely falling and accelerated radial observers cannot enter the black hole, since their radial velocity becomes zero at $r=2M$ and attains imaginary values for $r<2M$. This result indicates that, on the one hand, in the Lorentzian-Euclidean Schwarzschild black hole, the central singularity located at $r=0$ can be averted, and, on the other,   that   matter accumulates around its event horizon rather than being swallowed. Remarkably, the main observed features, such as the formation of accretion disks, remain unaltered,  since the outer geometry is the same as the ordinary Schwarzschild one, with differences arising only in the near-horizon region.

The Lorentzian-Euclidean Schwarzschild metric can be constructed by assuming a shift in the nature of  time variable $t$, which passes from being real-valued to imaginary as the event horizon is traversed. We  proposed to interpret this mechanism through the concept of {\it atemporality} \cite{Capozziello2024}, which thus acts, in our context, as the dynamical means  to \emph{avoid} the black-hole singularity. By demonstrating that the regularized Kretschmann invariant is always bounded along  radial trajectories, we  interpreted this scalar quantity  as the key indicator of the atemporality structure.

In this paper, we further explore the properties of the Lorentzian-Euclidean black hole. The work is organized as follows. We briefly outline the main aspects of the Lorentzian-Euclidean Schwarzschild geometry in  Sec. \ref{Sec:Lorentzian-Euclidean-BH}. Then,  we work out a  set of signature-adaptive local coordinate transformations  suited for our model in Sec. \ref{Sec:generalized-transformation}. This analysis prepares the grounds for the examination of null geodesics,  which will be carried out in Sec. \ref{Sec:null-geodesics}. As we will see,  light paths are also unable to cross the event  horizon, further supporting the conclusion that singularities can be avoided in our black hole model. The causal structure, including the underlying Kruskal, Finkelstein, and Penrose diagrams, is studied in Sec. \ref{Sec:Penrose-diagrams}. Subsequently, in Sec. \ref{Sec:Accretion-Matter}, we study the process of accretion of perfect fluids, which, as mentioned earlier, represents  one of the observational evidences for the existence of black holes.  Finally, we  draw  conclusions in Sec. \ref{Sec:Conclusion}. 

Throughout the paper, we use  units $G=c=1$. In addition, Greek indices $\alpha,\beta,\dots=0,\dots,3$  are coordinate indices, whereas Latin ones $a,b,\dots=\hat{0},\dots,\hat{3}$    are tetrad indices.

\section{The Lorentzian-Euclidean black hole}\label{Sec:Lorentzian-Euclidean-BH}

In this section, we briefly recall the main aspects of the Lorentzian-Euclidean black hole set up in Ref. \cite{Capozziello2024}. After  outlining  its geometrical properties in Sec. \ref{Sec:Geometry}, in Sec. \ref{Sec:Radially-infall-particles} we deal with the motion of radially infalling bodies, which reveals how the $r=0$ singularity can be averted  and the ensuing notion of  atemporality emerges.

\subsection{The geometry}\label{Sec:Geometry}

The Lorentzian-Euclidean black-hole geometry features the presence of a signature-changing metric which becomes degenerate on the event horizon. The spacetime manifold can be indeed divided into two regions: the Lorentzian domain $\mathscr{D}_+$ and the Euclidean zone $\mathscr{D}_-$.  In passing from the former region, defined by the condition $r>2M$, to the latter, where $r<2M$, the signature changes from the standard  Lorentzian to an ultrahyperbolic one. In particular, in $\mathscr{D}_-$ the metric acquires the same features as the Euclidean Schwarzschild metric\footnote{It should be noted that the standard Euclidean Schwarzschild metric fulfills the condition $r \geq 2M$ in the Euclidean section of the complexified Schwarzschild spacetime; see e.g. Refs. \cite{GH1977,Esposito1994,Battista-Esposito2022,Garnier2025} for further details.}, thus  explaining the name  we gave to this solution. Upon adopting spherical coordinates $x^\mu= (t,r,\theta,\phi)$, the metric can be expressed as  \cite{Capozziello2024}
\begin{align}
\dd s^2 &=g_{\mu \nu} \dd x^\mu \dd x^\nu
\nonumber \\    
    &= - \varepsilon \left( 1-\frac{2M}{r} \right) {\rm d} t^2  
+ \dfrac{{\rm d} r^2 }{\left(1-2M/r\right)} 
+ r^2 \dd \Omega^2,
\label{Lorentzian-Euclidean-Schwarzschild}
\end{align}
where  $\dd \Omega^2 =  {\rm d} \theta^2  + \sin^2 \theta \; 
{\rm d} \phi^2 $ and
\begin{widetext}
\begin{align}
\varepsilon = {\rm sign} \left( 1 - \frac{2M}{r}\right)= 2 H \left( 1 - \frac{2M}{r}\right)-1 =
\left \{ \begin{array}{rl}
& 1, \qquad \;\; \, {\rm if}\; r>2M,\\
& 0,  \qquad \;\; \, {\rm if}\;  r=2M,\\
& -1, \qquad {\rm if}\;  r<2M,\\
\end{array}
\right.,
\label{epsilon-of-r}
\end{align}
\end{widetext}
with $H\left(1-2M/r\right)$  the step function with normalization  $H(0)=1/2$. 

It is clear that, in these coordinates, the metric \eqref{Lorentzian-Euclidean-Schwarzschild} appears to be both divergent and degenerate on the event horizon. The first issue can be easily sorted out by resorting to a different coordinate system, and  in Ref. \cite{Capozziello2024} we  chose the  Gullstrand-Painlev\'e coordinates.  In this way, by exploiting the Hadamard \emph{partie finie} regularization method \cite{Blanchet2000}, we were able to regularize the Dirac-delta-like singularities occurring in the Riemann and Weyl  tensors, which stem from the presence of the step function in Eq. \eqref{epsilon-of-r}. This process  involved the use of   some methods usually applied to the analysis of gravitational shock-waves  (see e.g. Refs. \cite{Dray1984,Sfetsos1995,Battista_Riemann_boosted}) and the introduction of a smooth expression for the sign function \eqref{epsilon-of-r}, whose behavior near $r=2M$ has been approximated with 
\begin{align}
\varepsilon(r) = \frac{\left(r-2M\right)^{1/(2 \kappa +1)}}{\left[\left(r-2M\right)^2 + \varrho\right]^{1/2(2 \kappa +1)}}, 
\label{sign-regularization}
\end{align}
where $\varrho$ is a small positive quantity having the dimensions of a squared length, and $\kappa $ a large positive integer.  By means of this regularization procedure, we  proved, on the one hand, that  $r=2M$ is not the site of a surface layer or an impulsive gravitational wave, and, on the other, that Eq. \eqref{Lorentzian-Euclidean-Schwarzschild} is a well-defined signature-changing solution of  vacuum Einstein equations \cite{Capozziello2024}.

The Euclidean region $\mathscr{D}_-$ can be equivalently described as the domain where the  time variable  attains imaginary values.   We  proposed to associate this feature with the emergence of atemporality. This concept assumes a dynamical nature in our setup, as it configures as the mechanism  which permits  avoiding the black-hole singularity located at $r=0$. This point will be  briefly described in the next section. Further details can be found in Ref. \cite{Capozziello2024}.

\subsection{The radially infalling observer and the  avoidance of  singularity}\label{Sec:Radially-infall-particles}

In the spacetime geometry \eqref{Lorentzian-Euclidean-Schwarzschild},  the geodesic equation is related,  as usual, to the vanishing of the four-acceleration vector:
\begin{align}
a^\mu = u^\lambda \nabla_\lambda u^\mu=0.
\label{four-acceleration}
\end{align}
Unlike the standard scenario, in our framework the four-velocity  
\begin{align}
u^\mu = \frac{\dd x^\mu}{  \dd \sigma}:= \dot{x}^\mu,
\label{u-mu-d-sigma} 
\end{align}
satisfies the generalized  normalization condition
\begin{align}
g_{\mu \nu} u^\mu u^\nu = - \varepsilon,
\label{velocity-norm}
\end{align}
with $\sigma$  an affine parameter  which  corresponds to the proper time $\tau = \int \sqrt{-\dd s^2}$ in the Lorentzian regime $\mathscr{D}_+$ and it is constructed in such a way to be continuous across the event horizon (see Refs. \cite{Ellis1992a,Ellis1992b} for further details). It  is easy to show that the above equations, evaluated in   the equatorial plane $\theta=\pi/2$, yield \cite{Capozziello2024}
\begin{align}
\dot{r}^2 + \left(1-\frac{2M}{r}\right) \left(\frac{L^2}{r^2} + \varepsilon \right)= \frac{E^2}{\varepsilon^3},
\label{geodesic-equation}
\end{align}
where 
\begin{align}
E&= -\varepsilon g_{\mu \nu} \xi^\mu u^\nu=\varepsilon^2 \left(1-\frac{2M}{r}\right) \dot{t},
\label{energy-PG}
\\
L&=g_{\mu \nu} \psi^\mu u^\nu= r^2 \dot{\phi},
\label{angular-momentum-PG}
\end{align}
are the constants of motion associated with the static Killing vector $\xi^\mu = \delta^{\mu t}$ and the rotational one $\psi^\mu= \delta^{\mu \phi}$, respectively. Notice the presence of the $\varepsilon$ factor in Eq. \eqref{energy-PG}, which reflects the structure of Eq. \eqref{velocity-norm}. Physically, $E$ and $L$ represent the total conserved energy (including the potential energy due to the gravitational field) and angular momentum per unit rest mass of the freely falling body,  respectively \cite{Carroll2004}.

It follows,  from Eqs. \eqref{geodesic-equation}--\eqref{angular-momentum-PG}, that the four-velocity of the freely falling   observer  radially approaching the Lorentzian-Euclidean black hole in the equatorial plane is
\begin{align}
u^\mu=   \left[\frac{E}{\varepsilon^2} \left(\frac{1}{1-2M/r}\right),-\sqrt{\varepsilon}\sqrt{\frac{2M}{r}-1+\frac{E^2}{\varepsilon^4}},0,0\right],
\label{free-fall-velocity}
\end{align}
which boils down to
\begin{align}
\tilde u^\mu=  \left[ \frac{1}{1-2M/r},-\sqrt{\varepsilon}\sqrt{\frac{2M}{r}},0,0\right], 
\label{free-fall-velocity-infinity}
\end{align}
if the observer has $E=\varepsilon^2$, i.e., the motion  begins  at rest from infinity. 

The radial velocity of the geodetic bodies vanishes at the event horizon and becomes imaginary inside the black hole. This is evident from Eq. \eqref{free-fall-velocity-infinity}, and can be inferred from Eq. \eqref{free-fall-velocity} by recalling that Eq. \eqref{energy-PG} implies 
\begin{align}
 E=\alpha \varepsilon^2,
 \label{E-alpha-varepsilon-2}
\end{align}
$\alpha$ being some positive-definite, bounded function. This serves as an indication that the black-hole singularity, located at $r=0$, can be avoided. 

As detailed in Ref. \cite{Capozziello2024}, an additional novel facet of the Lorentzian-Euclidean pattern is that the behavior of infalling objects is  identical  from both the particle perspective and that of an observer stationed at infinity.  Indeed, like in the standard Schwarzschild solution, the coordinate time $t$ attains unboundedly large values when $r=2M$; however, in contrast to the standard scenario, a similar behavior occurs also in the  proper reference frame of the body, as  an infinite amount of proper time $\sigma$ is required to reach the event horizon.   

In our model, atemporality assumes a dynamical role by allowing  a cutoff for the radial trajectory $r(\sigma)$, which  enforces   the constraint $2M \leq r(\sigma) <\infty$. Consequently, the Kretschmann  invariant $\mathcal{K}=R_{\alpha \beta \mu \nu} R^{\alpha \beta \mu \nu}$ is bounded along the path of radially moving observers, and can be interpreted as the parameter related to the implementation of the atemporality mechanism.  We will come back to the concept of atemporality in Sec. \ref{Sec:radial-photons}, where we will deal with radial null rays.

\section{A generalized set of coordinate transformations}\label{Sec:generalized-transformation}




In Einstein gravity, the Equivalence Principle enables the construction of a set of four vector fields, called tetrads,   
which permit  recovering the local Minkowski structure   of   spacetime manifold at any point \cite{Weinberg1972,DeSabbata1986,Nakahara2003,Hartle2003,Carroll2004,Gasperini2013,Aldrovandi2013,Ortin2015}. These tetrads serve as a noncoordinate basis for the local tangent space, which is equipped with a Minkowski metric tensor. Moreover, local Lorentz invariance requires  that both the field equations and the action remain invariant under pointwise Lorentz transformations relating the tetrads at each point. Among the various possibilities afforded by this freedom, a notable one is that the proper reference frames of two distinct observers can be connected at each point via a Lorentz boost, thereby allowing a comparison between their respective measurements (see Refs. \cite{MTW,Morris1988} for the general formalism and Ref. \cite{Battista2020} for an application to cosmological settings). 

In our context, the situation is different owing to the peculiarities exhibited  by the geometry \eqref{Lorentzian-Euclidean-Schwarzschild}. Consequently, the main aspects characterizing  the tetrad fields, such as their transformation properties in the tangent space and  orthogonality condition, should account for the signature-changing nature and the degeneracy of the Lorentzian-Euclidean Schwarzschild metric. For this reason, in this section, we generalize the local Lorentz transformations, employed in General Relativity, by working out a set of signature-adaptive changes of coordinates  suited for our model, which we dub  \qm{Lorentzian-Euclidean} transformations. Starting from the analysis of their global form in Sec. \ref{Sec:Lorentz-and-rotation}, their local version will then be presented  in Sec. \ref{Sec:Local-Lorentzian-Euclidean-transf}. As we will see, this approach  provides  further insights into the  features of our geometry and proves useful for  investigating  the dynamics of massless particles in Sec. \ref{Sec:null-geodesics}.

\subsection{Global   Lorentzian-Euclidean transformations in flat space }\label{Sec:Lorentz-and-rotation}

It is well-known that any two tetrads at a given point are equivalent if they uphold their orthonormality condition   \cite{Carroll2004}. This means that tetrads can be locally related by either  Lorentz  or  orthogonal transformations, which represent the changes that preserve the  flat tangent-space metric, depending on whether the spacetime metric signature is Lorentzian or Euclidean. Therefore, in view of the discussion in Sec. \ref{Sec:Local-Lorentzian-Euclidean-transf}, we first need to devise a global set of signature-adaptive coordinate changes encompassing both  Lorentz transformations and ordinary Euclidean rotations. 

Lorentz boosts are global coordinate transformations connecting two inertial frames moving with relative velocity $V$ in flat Minkowski space \cite{MTW}. If the relative motion is along the common $x$ axis, these transformations can be written as
\begin{subequations}
\label{Lorentz-transf-flat-1}
\begin{align}
t^\prime &= \gamma_{(1)} \left(t-V x\right), 
\\
x^\prime &= \gamma_{(1)} \left(x-V t\right), 
\label{x-prime-Lorentz-1}
\end{align}
\end{subequations}
jointly with $y^\prime=y$ and $z^\prime=z$, where
\begin{align}
\gamma_{(1)} := \frac{1}{\sqrt{1-V^2}}.   
\label{gamma-plus-1}
\end{align}
These relations can be  parametrized in terms of the variable $\beta$, defined by the identity $V=\tanh \beta$, by letting
\begin{align}
\cosh \beta= \gamma_{(1)}, \qquad \sinh \beta = V \gamma_{(1)},  
\end{align}
and hence we obtain
\begin{subequations}
\label{Lorentz-transf-flat-2}
\begin{align}
t^\prime &= t \cosh \beta - x \sinh \beta, 
\\
x^\prime &= - t \sinh \beta + x \cosh \beta.
\label{x-prime-Lorentz-2}
\end{align}
\end{subequations}

Suppose now to deal with a rotation by an angle $\theta$ in the  $t-x$ plane. Although there is no  clear physical interpretation for the rotation of  time variable, in this case Eq. \eqref{Lorentz-transf-flat-2} should be replaced by the orthogonal transformation
\begin{subequations}
\label{rotation-flat-1}
\begin{align}
t^\prime &= t \cos \theta - x \sin \theta, 
\\
x^\prime &=  t \sin \theta + x \cos \theta.
\label{x-prime-rotation-1}
\end{align}
\end{subequations}
Similarly, we can define the factor 
\begin{align}
 \gamma_{(-1)} := \frac{1}{\sqrt{1+V^2}},  
 \label{gamma-minus-1}
\end{align}
which leads  to the identifications   
\begin{align}
\cos \theta = \gamma_{(-1)}, \qquad \sin \theta = V \gamma_{(-1)},
\end{align}
with $V=\tan \theta$. In this way, Eq. \eqref{rotation-flat-1} amount to
\begin{subequations}
\label{rotation-flat-2}
\begin{align}
t^\prime &= \gamma_{(-1)} \left(t-V x\right), 
\\
x^\prime &= \gamma_{(-1)} \left(x+V t\right). 
\label{x-prime-rotation-2}
\end{align}
\end{subequations}
A comparison between Eqs. \eqref{gamma-plus-1} and \eqref{gamma-minus-1} shows that $\gamma_{(1)}$ and $\gamma_{(-1)}$ differ by the sign in front of  $V^2$; furthermore, we see that there is a sign difference involving the second factor in the right-hand side of  Eqs.  \eqref{x-prime-Lorentz-1} and \eqref{x-prime-rotation-2} (equivalently, we have that, apart from the different nature of the  functions involved, there is a sign difference in the first factor of the right-hand side of Eqs. \eqref{x-prime-Lorentz-2} and \eqref{x-prime-rotation-1}). 

Given these premises, let us consider the flat-space signature-changing metric
\begin{align}
\dd s^2_{\rm flat}= f_{ab} \dd x^a \dd x^b= -\varepsilon \dd t^2 + \dd x^2 + \dd y^2 + \dd z^2, 
\label{flat-metric}        
\end{align}
where $\varepsilon=0,\pm 1$.  This metric becomes degenerate when $\varepsilon=0$,  allowing us to view the underlying spacetime as consisting of two disjoint flat manifolds, with $f_{ab}$ taking either a Lorentzian or a Euclidean signature depending on whether  $\varepsilon=1$ or $\varepsilon=-1$. We might thus define the  isometries of this space by combining the  two sets of transformations \eqref{Lorentz-transf-flat-1} and \eqref{rotation-flat-2}  into one, as follows:
\begin{align}
t^\prime &= \hat \gamma \left(t-V x\right), 
\nonumber \\
x^\prime &= \hat \gamma \left(x- \varepsilon V t\right), 
\label{general-transformations-trial}
\end{align}
with 
\begin{align}
\hat{\gamma }:= \frac{1}{\sqrt{1-\varepsilon V^2}}.
\label{hat-gamma}
\end{align}
The factor $\varepsilon$ occurring above allows for the sign flip distinguishing Eqs. \eqref{Lorentz-transf-flat-1} and \eqref{rotation-flat-2}, and Eqs. \eqref{gamma-plus-1} and \eqref{gamma-minus-1}.  However, Eqs. \eqref{general-transformations-trial} and \eqref{hat-gamma} do not represent faithfully the isometries of the geometry \eqref{flat-metric}, since they fail to take into account the degenerate status of the metric $f_{ab}$ at $\varepsilon=0$. To bring this key aspect into consideration, let us substitute Eq. \eqref{hat-gamma}  with the parameter
\begin{align}
\gamma := \frac{1}{\sqrt{1-V^2/\varepsilon}},   
\label{gamma-flat-space} 
\end{align}
which boils down to $\gamma_{(1)}$ or $\gamma_{(-1)}$ when $\varepsilon=1$ or $\varepsilon=-1$, respectively. Since we are assuming a finite $V$, $\gamma$ vanishes as soon as $\varepsilon=0$. Following this line of reasoning, let us then replace Eq. \eqref{general-transformations-trial} with the formulas
\begin{subequations}
\label{general-transformations}
\begin{align}
t^\prime &=  \gamma \left(t-V x\right), 
\\
x^\prime &= \gamma \left(x- \frac{V}{\varepsilon} t\right), 
\label{general-transformation-x-prime}
\end{align}
\end{subequations}
which reflect the  degeneracy of $f_{ab}$, since they completely lose their meaning when $\varepsilon=0$. In other words,  by construction,  transformations \eqref{general-transformations} make sense only if $\varepsilon \neq 0$. This point can also be realized by noting that relations \eqref{general-transformations} leave invariant the combination 
\begin{align}
-\frac{t^2}{\varepsilon} + x^2 +y^2+z^2,
\end{align}
which is ill-defined if $\varepsilon=0$, while it amounts to the  Lorentzian/Euclidean spacetime invariant
\begin{align}
-\varepsilon t^{ 2} + x^{ 2} +y^{ 2}  +z^{ 2}, 
\label{interval}
\end{align}
only if $\varepsilon = \pm 1$. Equivalently, we have,  from Eq. \eqref{general-transformations},
\begin{align}
& -\varepsilon t^{\prime 2} + x^{\prime 2} +y^{\prime 2}  +z^{\prime 2} = \gamma^2 \left[-\varepsilon t^2 \left(1-V^2/\varepsilon^3 \right) \right.
\nonumber \\
& \left. + x^2 \left(1-\varepsilon V^2 \right) + 2 Vtx \left(\varepsilon-\varepsilon^{-1}\right)\right] +y^2+z^2,
\end{align}
which yields $\mp  t^{ 2} + x^{ 2} +y^{ 2}  +z^{ 2}$ when  $\varepsilon= \pm 1$. 

From the above analysis it is clear that Eq. \eqref{general-transformations} gives rise to two detached sets of transformations suitable for the disjoint union of  flat Minkowski and Euclidean four-dimensional spaces. This set of signature-adaptive coordinate changes thus represent the global hybrid Lorentzian-Euclidean transformations we pursued. As a final remark, it should be noticed that the quantity $V$, occurring in Eq. \eqref{general-transformations}, is left completely arbitrary. However, this is not  a hindrance, as the procedure for evaluating it locally will be clear in the next section (see Eq. \eqref{V-parameter-local}, below).

\subsection{Local   Lorentzian-Euclidean transformations  } \label{Sec:Local-Lorentzian-Euclidean-transf}

Let us determine now the pointwise version of the hybrid Lorentzian-Euclidean transformations starting from their previously established global form \eqref{general-transformations}. 

We begin  by briefly setting forth the  tetrad formalism  in Sec. \ref{Sec:Tetrad-formalism}. Subsequently, the explicit expression of our local signature-adaptive transformations will be  shown in Sec. \ref{Sec:static-freely-frames}, where we construct the reference frames of a static body and a freely falling particle, assuming that they meet at some radial distance $r$  outside the event horizon. The underlying formulas will then be applied in Sec. \ref{Sec:energy-measured-static} to evaluate the energy of the geodesic particle locally measured by the observer at rest. 

The tools described in this section will prove useful  for both    the study of null geodesics and  matter accretion (see Secs. \ref{Sec:null-geodesics} and \ref{Sec:Accretion-Matter}).

\subsubsection{The tetrad formalism}\label{Sec:Tetrad-formalism}

The geometric description of gravity can be achieved via two different but equivalent paradigms.  The first  relies on the  language of differential geometry and the subsequent choice of coordinate bases in the tangent space. The second approach is based   on noncoordinate bases of the tangent space, referred to as tetrads or {\it vierbeins} \cite{Gasperini2013}. This means that  any vector $\boldsymbol{\xi}$ can be expressed in the noncoordinate basis $\{ \boldsymbol{e}_a \}$ as
\begin{align}
  \boldsymbol{\xi} = \xi^a   \boldsymbol{e}_a, 
\end{align}
the flat tangent-space components $\xi^a$  being related to the coordinate-basis ones $\xi^\mu$ by
\begin{align}
\xi^a = e^a{}_{\mu}(x) \xi^\mu, 
\end{align}
where $\{e^a{}_{\mu}(x)\}$ form a point-dependent, invertible four-dimensional matrix. In the same vein, the tetrad $\{\boldsymbol{e}_a \}$ can be  expressed in terms of the ordinary coordinate basis $\{\boldsymbol{e}_\mu \}$ associated with the coordinate system $x^\mu$ through
\begin{align} \label{tetrad-formula-1}
\boldsymbol{e}_a =e_a{}^{\mu}  (x) \boldsymbol{e}_\mu,  
\end{align}
with $e_a{}^{\mu}  (x)$ the inverse of $e^a{}_{\mu}(x)$. 

In the Einstein theory, tetrads span the local Minkowski flat space tangent to the spacetime manifold at the given point, where, according to the  Equivalence Principle,  physical laws are Lorentz invariant  \cite{DeSabbata1986}. For this reason, the basis   $\{\boldsymbol{e}_a\}$ can be chosen in such a way to be orthonormal with respect to the spacetime metric $g_{\mu \nu}$. In our Lorentzian-Euclidean framework, we can generalize this picture by supposing that $\{ \boldsymbol{e}_a \}$ locally generate either a  Minkowski space or a Euclidean space, and hence we require that
\begin{align}
 e_a{}^{\mu} e_b{}^{\nu}  g_{\mu \nu} = f_{ab}, 
 \label{orthonormality-tetrad-1}
\end{align}
or, equivalently,
\begin{align}
g_{\mu \nu} = e^a{}_\mu e^b{}_\nu f_{ab},
\label{g-tetrad-metric-f}
\end{align}
where  $f_{ab}$ is the signature-changing flat metric introduced in Eq. \eqref{flat-metric}. The  last relation  implies 
\begin{align}
g=e^2 (-\varepsilon),
\end{align}
where $g:= \det g_{\mu \nu}$ and $e:= \det e^a{}_\mu$, thus taking into account the degeneracy of  $g_{\mu \nu}$.

The local  coefficients $\{ e^a{}_\mu(x)  \}$ fully determines the spacetime metric $g_{\mu \nu}(x)$ at any given point up to a residual freedom arising from the choice of the tetrads, owing to the transformations $\Lambda^a{}_b(x)$ in the local tangent space. These act on the tangent-space coordinates $x^a$  as $x^{\prime a} = \Lambda^{a}{}_b x^b$ and comply with the $\varepsilon$-dependent relation
\begin{align} \label{Lambda-property}
    f_{cd} \Lambda^{c}{}_a \Lambda^{d}{}_b = f_{ab}.
\end{align}
Consequently, both $ e^a{}_\mu  $ and the transformed 
\begin{align} \label{transformed-tetrad}
{e^\prime}^{a}{}_\mu  = \Lambda^{a}{}_b  e^b{}_\mu  
\end{align}
yield the same metric,  as
\begin{align}
g^\prime_{\mu \nu}&= {e^\prime}^{a}{}_\mu {e^\prime}^{b}{}_\nu f_{ab} = \Lambda^{a}{}_c \Lambda^{b}{}_d {e}^{c}{}_\mu {e}^{d}{}_\nu f_{ab} 
  \nonumber \\  
& = f_{cd} {e}^{c}{}_\mu {e}^{d}{}_\nu \equiv g_{\mu \nu},
\end{align}
where we have taken into account Eq. \eqref{Lambda-property}.

Since the local frame transformations $\Lambda^a{}_b(x)$ preserve the flat metric $f_{ab}$, they embody the sought-after local hybrid Lorentzian-Euclidean transformations. Their explicit form will be displayed in next section (see Eq. \eqref{general-transformations-curved}, below). 

\subsubsection{The frames of the static and freely falling observers}\label{Sec:static-freely-frames}

We now provide a physical realization of tetrad fields introduced earlier and present the explicit expression of the local transformations $\Lambda^a{}_b(x)$ that relate them.

In the class of accelerated observers, an important role is fulfilled by the static ones. Bearing in mind the generalized normalization condition \eqref{u-mu-d-sigma}, their four-velocity $u^\mu_{\rm s}$ can be expressed in the standard coordinates $x^\mu= (t,r,\theta,\phi)$  as
\begin{align}
u^\mu_{\rm s}=  \left[\frac{1}{\sqrt{1-2M/r}},0,0,0\right].
\label{static-oberver-velocity}
\end{align}

At this stage,  we can resort to the formalism put forth in the previous section to construct the proper reference frame  for the  observer at rest. In fact, the (noncoordinate) basis $\{\boldsymbol{e}_a\}$,  carried by this  observer, is defined by means of the linear combination \eqref{tetrad-formula-1} and hence takes the form  
\begin{subequations}
\label{static-reference-frame}
\begin{align}
\boldsymbol{e}_{\hat t}&= e_{\hat t}{}^{t}  \boldsymbol{e}_t =\frac{1}{\sqrt{1-2M/r}}\boldsymbol{e}_t = \boldsymbol{u}_{\rm s},
\label{e-hat-t}\\
\boldsymbol{e}_{\hat r}&= e_{\hat r}{}^{r}  \boldsymbol{e}_r= \sqrt{1-2M/r} \, \boldsymbol{e}_r,
\\
\boldsymbol{e}_{\hat \theta}&= e_{\hat \theta}{}^{\theta}  \boldsymbol{e}_\theta= \frac{1}{r} \boldsymbol{e}_\theta,
\\
\boldsymbol{e}_{\hat \phi}&= e_{\hat \phi}{}^{\phi}  \boldsymbol{e}_\phi= \frac{1}{r \sin \theta} \boldsymbol{e}_\phi,
\end{align}    
\end{subequations}
which immediately yields $\boldsymbol{e}_{a} \cdot \boldsymbol{e}_{b}={\rm diag}(-\varepsilon,1,1,1)=f_{ab}$ (cf. Eq. \eqref{orthonormality-tetrad-1}). 

Having determined the set $\{\boldsymbol{e}_a\}$, we can now construct the proper reference frame, denoted by $\{\boldsymbol{e}_{a^\prime} \}$, parallel transported by the geodetic observer. Both the recipe of Sec. \ref{Sec:Lorentz-and-rotation} and the tetrad transformation properties discussed in Sec. \ref{Sec:Tetrad-formalism} will be essential. First of all, we need to calculate the  Lorentz factor $\gamma$ of the transformation connecting the two frames. Drawing from Eq. \eqref{gamma-flat-space}, we can write 
\begin{align}
 \gamma=\frac{1}{\sqrt{1-\mathscr{V}^2(r)/\varepsilon}}, 
 \label{gamma-factor}
\end{align}
where the boost parameter $\mathscr{V}(r)$ amounts to the radial velocity of the freely falling particle locally measured by the static observer at the radius $r$ where the two bodies meet (the notation has been chosen to stress the difference with the globally-defined $V$ appearing in the formulas of Sec. \ref{Sec:Lorentz-and-rotation}). Similarly to Special Relativity, $\mathscr{V}(r)$ reads as
\begin{widetext}
\begin{align}
\mathscr{V}(r)&=   \frac{{\rm proper \; radial \; distance \; as \; measured \; by \; the \;  static \; observer}}{{\rm proper \; lapse \; time \; as \; seen \; by \; the \;  static \; observer}} =\frac{\sqrt{g_{rr}}}{\sqrt{-\dfrac{g_{tt}}{\varepsilon}} } \frac{\dd r}{\dd t},
\label{V-parameter-local}
\end{align}
\end{widetext}
which, upon using Eqs. \eqref{Lorentzian-Euclidean-Schwarzschild} and \eqref{free-fall-velocity}, yields
\begin{align}
\mathscr{V}(r) &= \left(1-2M/r\right)^{-1} \frac{\dd r}{\dd t}=  \left(1-2M/r\right)^{-1} \frac{\dd r}{\dd \sigma} \left(\frac{\dd t}{\dd \sigma}\right)^{-1}
\nonumber \\  
&= -\sqrt{\varepsilon} \, \frac{\varepsilon^2}{E} \sqrt{\frac{2M}{r}-1+\frac{E^2}{\varepsilon^4}},
\label{boost-velocity-explicit}
\end{align}
and consequently, from Eq. \eqref{gamma-factor}, we have
\begin{align}
\gamma= \frac{E}{\varepsilon^2} \left(\frac{1}{\sqrt{1-2M/r}}\right).   
\label{gamma-explicit} 
\end{align}

The orthonormal basis $ \{ \boldsymbol{e}_{a^\prime} \} $ of the freely falling particle can now be obtained by applying a  boost\footnote{This is actually a Lorentz boost, since,  by hypothesis, the two bodies meet at $r>2M$. However, the obtained formulas \eqref{general-transformations-curved} are formally valid also for $r<2M$. }  along the $\boldsymbol{e}_{\hat r} $ direction to the stationary observer frame $ \{ \boldsymbol{e}_{a} \} $  (cf.  Eq. \eqref{static-reference-frame}),  with parameters  $ \mathscr{V}(r)$ and $\gamma$ given above.  This coordinate change represents the local version of the global transformations displayed in Eq. \eqref{general-transformations}, and hence  it is given by
\begin{subequations}
\label{general-transformations-curved}
\begin{align}
\boldsymbol{e}_{\hat{0}^\prime} &= \gamma \left(\boldsymbol{e}_{\hat t} +  \mathscr{V}(r) 
\boldsymbol{e}_{\hat r} \right) =\boldsymbol{u},
\label{e-hat-zero-prime} \\
\boldsymbol{e}_{\hat{1}^\prime}   &= \gamma \left(\boldsymbol{e}_{\hat r} +  \frac{\mathscr{V}(r)}{\varepsilon } \boldsymbol{e}_{\hat t} \right), 
\label{e-hat-1-prime} \\
\boldsymbol{e}_{\hat{2}^\prime}   &= \boldsymbol{e}_{\hat \theta},
\\
\boldsymbol{e}_{\hat{3}^\prime}   &= \boldsymbol{e}_{\hat \phi},
\end{align}
\end{subequations}
where $\boldsymbol{u}$ can be read off from  Eq. \eqref{free-fall-velocity} and the $\varepsilon$ factor appearing in Eq. \eqref{e-hat-1-prime}  reflects the $\varepsilon$-dependent correction in Eq. \eqref{general-transformation-x-prime}.  It is easy to show that 
\begin{align}
\boldsymbol{e}_{a^\prime} \cdot \boldsymbol{e}_{b^\prime}={\rm diag}(-\varepsilon,1,1,1),    
\end{align}
where, in particular, the identity $\boldsymbol{e}_{\hat{0}^\prime} \cdot \boldsymbol{e}_{\hat{0}^\prime} = -\varepsilon$ easily follows from Eq. \eqref{velocity-norm}. The above relation represents a crucial consistency check in support of  the correctness of the arguments of Sec. \ref{Sec:Lorentz-and-rotation}, and, specifically, of  the form  \eqref{general-transformations} assumed by the global Lorentzian-Euclidean transformations. Moreover, it demonstrates that the orthonormality condition \eqref{orthonormality-tetrad-1} has been preserved,  thereby confirming that Eq. \eqref{general-transformations-curved} furnishes an explicit realization of local tangent space transformations $\Lambda^a{}_b(x)$ (cf. Eq. \eqref{transformed-tetrad}).

A few remarks are now necessary. The generalized coordinate changes  devised here are suited for a metric exhibiting both a Euclidean and Lorentzian signature, while the Schwarzschild metric \eqref{Lorentzian-Euclidean-Schwarzschild} admits an ultrahyperbolic signature inside the black hole. However, the use of  formulas \eqref{general-transformations-curved} is justified by the fact that the ultrahyperbolic nature of $g_{\mu \nu}$ arises from its  Euclidean structure  in the region $r<2M$. Another crucial observation is the following. One may notice that the boost velocity \eqref{boost-velocity-explicit}  is zero at the event horizon (where the velocity of both  static and free-fall observers vanishes), unlike the ratio $\mathscr{V}^2(r)/\varepsilon$ entering  the definition \eqref{gamma-factor}. This point suggests that the use of a Lorentz factor defined, for $r>2M$, as $\bar{\gamma} := 1/\sqrt{1-\mathscr{V}^2(r)}$, would have yielded well-defined Lorentz transformations in the limit $r \to \left(2M\right)^+$ (this option  corresponds to the choice considered in Eqs. \eqref{general-transformations-trial} and \eqref{hat-gamma}). However, this conceivable set of transformations does not accommodate the degenerate character of the Lorentzian-Euclidean Schwarzschild metric;  on the other hand, transformations \eqref{general-transformations-curved} involving the Lorentz parameter \eqref{gamma-factor} have the advantage of becoming ill-defined  at the event horizon. The same situation  also occurs in  the ordinary Schwarzschild spacetime \cite{MTW}, but, in our scenario, it can be interpreted as a consequence of  the violation of  Equivalence Principle at those points where the metric is  degenerate  (recall indeed that the statement $\det g_{\mu \nu}=0$ is coordinate invariant; see e.g. Ref. \cite{Gunther} for details). In addition, the fact that   Eq. \eqref{gamma-explicit} attains the same form, modulo the $\varepsilon^2$ contribution, as in  Einstein gravity,  ties in with  the spirit of our model, where we have constructed a black hole metric  assuming analogous characteristic as  the standard Schwarzschild solution outside the event horizon. Lastly, let us stress that the transformations \eqref{general-transformations} and those employed in this section  are   inevitably diverse due to  their distinct nature: the former are global coordinate changes, whereas  the latter are  local. This explains the different behavior of  $\gamma$ factor in the two regimes, where it is found to vanish  when $\varepsilon=0$ according to Eq. \eqref{gamma-flat-space} (recall that $V$ is supposed to be finite), while  becomes infinite if $r=2M$, and imaginary for $r<2M$ in view of Eq. \eqref{gamma-explicit}. This does not represent an issue, since the main purpose of our approach is achieved in both scenarios, with the two sets of coordinate transformations being indeterminate as soon as the metric becomes degenerate.

\subsubsection{The energy measured by the static observer}\label{Sec:energy-measured-static}

The results of Sec. \ref{Sec:static-freely-frames} permit to compute the   energy of the radially moving body, as  locally measured by the  observer at rest in her/his proper reference frame $\{\boldsymbol{e}_{a}\}$. As the geodetic body flies past the static observer in its radially infalling orbit, the energy  measured is either
\begin{align}
E^{({\rm s})} &=-\varepsilon \boldsymbol{e}_{\hat{0}^\prime} \cdot \boldsymbol{e}_{\hat t} = -\varepsilon g_{\mu \nu} u^\mu u^\nu_{\rm s} = \frac{E}{\sqrt{1-2M/r}},  
\label{energy-E-s}
\end{align}
or
\begin{align}
\tilde E^{({\rm s})} &=-\varepsilon \tilde{\boldsymbol{u}} \cdot \boldsymbol{e}_{\hat t} = -\varepsilon g_{\mu \nu} \tilde u^\mu u^\nu_{\rm s}=  \frac{\varepsilon^2}{\sqrt{1-2M/r}},
\label{energy-tilde-E-s}
\end{align}
depending on whether the free-fall motion starts at rest from infinity; it is worth recalling  that $E^{({\rm s})}$ and $\tilde E^{({\rm s})}$ physically refer to the intrinsic energy coming from the inertia and the motion of the geodetic particle \cite{Carroll2004}. The above relations are in agreement with the local Lorentzian-Euclidean transformation \eqref{general-transformations-curved} relating  the frames $\{\boldsymbol{e}_{a}\}$ and $ \{ \boldsymbol{e}_{a^\prime} \} $, as we have
\begin{align}
E^{\rm (s)} &= \varepsilon^2 \gamma,
\label{energy-static}
\\
\tilde E^{\rm (s)} &=  \varepsilon^2 \left( \left. \gamma \right \vert_{E=\varepsilon^2}\right),
\label{tilde-energy-static}
\end{align}
where $\varepsilon^2$ amounts, in our model, to the rest energy (per unit rest mass) of the free-fall particle.

It is easy to show that both $E^{\rm (s)}$ and $ \tilde E^{({\rm s})}$ blow up at $r=2M$. Indeed, bearing in mind Eq. \eqref{energy-tilde-E-s}, we have 
\begin{align}
\lim_{r \to \left(2M\right)^+} \frac{\varepsilon^2}{\sqrt{1-2M/r}}= + \infty,
\end{align}
since $\varepsilon(r)=1$ for $r>2M$  (cf. Eq. \eqref{epsilon-of-r}). This means that  also $E^{({\rm s})} $ is infinite on the event horizon owing to Eq. \eqref{E-alpha-varepsilon-2}.   The same conclusions are valid  also if we use the approximation \eqref{sign-regularization} for $\varepsilon$. In this case,  Eq. \eqref{energy-E-s} gives
\begin{align}
 \tilde E^{({\rm s})} = \frac{\sqrt{r}}{\left[\left(r-2M\right)^2 + \varrho\right]^{1/(2 \kappa +1)}} \left(r-2M\right)^{(3-2\kappa)/2(2\kappa+1)},   
\end{align}
which becomes infinite for $r$ approaching $2M$ in the limit of large $\kappa$ (the regularization scheme briefly outlined in Sec. \ref{Sec:Geometry} only requires $\kappa \geq 1$ \cite{Capozziello2024}; however, the larger  the value of $\kappa$, the better the behavior of $\varepsilon$ at $r=2M$ is reproduced). The situation is similar to standard General Relativity (see e.g. Ref. \cite{Mazzitelli2020}), where the analogs of Eqs. \eqref{energy-E-s} and \eqref{energy-tilde-E-s}  diverge if the geodetic particle and the observer at rest meet at the event horizon. However, although the outcomes of the two theories are the same,  the underlying motivations are completely different. In the ordinary Schwarzschild spacetime, it is  the extreme time dilation experienced by the static observer that causes the  energy of the infalling particle to be infinitely blueshifted \cite{MTW}. On the other hand, in our model, the divergent character of $ E^{\rm (s)}$ and $\tilde E^{\rm (s)}$ is due to the degeneracy of the metric. The structure of the unified Lorentz-Euclidean transformations discussed before, jointly with a close inspection of Eqs.  \eqref{energy-static} and \eqref{tilde-energy-static}, facilitates the comprehension of this point. Indeed,  if we had  employed the aforementioned parameter $\bar{\gamma} := 1/\sqrt{1-\mathscr{V}^2(r)}$, then we would have obtained well-defined expressions of the energy also on the event horizon (the energy would have been zero in this case). However, as pointed out before, this choice of the Lorentz factor does not adapt to the degenerate status of the metric, whose imprint is, on the contrary, contained in our formulas for $ E^{\rm (s)}$ and $\tilde E^{\rm (s)}$. This is an example of how the Lorentzian-Euclidean geometry can reproduce a well-known feature of the  Schwarzschild solution by adopting  different mechanisms.

A  final remark is now in order. The standard interpretation of the constant of motion $E$ as the energy measured by a static observer at infinity can be recovered from Eq.  \eqref{energy-E-s} in the limit $r \to \infty$ \cite{MTW}. This shows that the presence of the factor $\varepsilon$  in the definition \eqref{energy-PG} of the conserved energy is crucial. As we will see in the next section, the situation might, in principle, be diverse when dealing with  massless objects. A crucial help will come from   the Lorentzian-Euclidean transformations discussed in this section.

\section{Null geodesics }\label{Sec:null-geodesics}

In this section, we analyze the behavior of photons within the Lorentzian-Euclidean black hole geometry. A crucial distinction from the case of massive particles should soon be pointed out. The equations of motion of massive bodies can be derived starting from  the $\varepsilon$-dependent normalization condition \eqref{velocity-norm}. This  allowed us to write the formula $E= -\varepsilon g_{\mu \nu} \xi^\mu u^\nu$ (see Eq. \eqref{energy-PG}), where, as indicated  previously,  the factor $\varepsilon$ is essential for interpreting $E$ as the (total) energy per unit rest mass of the geodetic particle measured by a static observer at infinity, thereby recovering the standard interpretation used in  Einstein theory. When dealing with null geodesics the situation is somewhat different as the photon four-velocity $k^\mu$ satisfies the relation $g_{\mu \nu} k^\mu k^\nu =0$, which,  unlike Eq. \eqref{velocity-norm}, does not explicitly depend  on $\varepsilon$. Therefore, \emph{a priori}, it is unclear whether the conserved photon energy $\Omega$  should mirror the form of $E$ and thus be expressed  as  
\begin{align}
   \Omega=  - \varepsilon g_{\mu \nu} \xi^\mu k^\nu,
\end{align}
or  whether it should take the more general form
\begin{align}
   \Omega=  -\F  g_{\mu \nu} \xi^\mu k^\nu.
   \label{Omega-1}
\end{align}
Here, we allow   the  $\F$ function to  remain as generic as possible by assuming the only  requirement that  Eq. \eqref{Omega-1} should represent the photon energy with the correct sign at least  in the outer Lorentzian domain.  We will see that the generalized coordinate transformations introduced in Sec. \ref{Sec:generalized-transformation} will provide a further constraint on $\F$. 

Before beginning  our analysis, a final comment is in order. In principle, one might expect no  interesting findings to emerge from the inspection of null curves in a Euclidean spacetime, since, by definition, the line element is positive-definite and hence the condition $\dd s^2=0$ yields the trivial solution where all  components of the photon velocity are zero.  However, the Lorentzian-Euclidean Schwarzschild metric admits an ultrahyperbolic signature inside the event horizon, which justifies the examination of null geodesics   in our setup. 

We explore  radial geodesics   in Sec. \ref{Sec:radial-photons},  and examine the $\F $ function by assessing the  photons frequency measured by the static and free-fall experimenters  in Sec. \ref{Sec:F-function}.

\subsection{Radially infalling photons}\label{Sec:radial-photons}


The dynamics of photons can be worked out by exploiting the condition $g_{\mu \nu} k^\mu k^\nu=0$, where the  four-velocity is expressed in standard coordinates as $k^\mu = (k^t,k^r,k^\theta,k^\phi)=\dd x^\mu / \dd \lambda$, $\lambda$ being the affine parameter. Thus, it easily follows from  Eq. \eqref{Lorentzian-Euclidean-Schwarzschild} that null orbits lying on the equatorial plane $\theta= \pi/2$ of the Lorentzian-Euclidean black hole are described by
\begin{align}
    -\varepsilon \left(1-\frac{2M}{r} \right) \left(k^t\right)^2 + \frac{\left(k^r\right)^2}{\left(1-2M/r \right)} +r^2 \left(k^\phi\right)^2=0. \label{null-geod-1}
\end{align}
The time independence and rotational symmetry of the solution \eqref{Lorentzian-Euclidean-Schwarzschild} imply the presence of two constants of motion. For the first quantity, we follow the choice \eqref{Omega-1} and write the general relation
\begin{align}
\Omega &=- \F g_{\mu \nu} \xi^\mu k^\nu= \F \varepsilon \left(1-\frac{2M}{r}\right) k^t,
\label{Omega-photon}
\end{align}
whereas for the second we adopt the usual formula
\begin{align}
\ell &= g_{\mu \nu} \psi^\mu k^\nu = r^2 k^\phi. 
\label{L-photon}
\end{align}
Equations \eqref{Omega-photon} and \eqref{L-photon} are the counterparts of Eqs. \eqref{energy-PG} and \eqref{angular-momentum-PG} in the particle case, as they represent the photon energy and angular momentum, respectively.  

By substituting the expressions for $\Omega$ and $\ell$ in Eq. \eqref{null-geod-1}, the photons equation of motion  assumes the simplified form
\begin{align}
    \left(k^r\right)^2 = \frac{\Omega^2}{\F^2 \varepsilon} -\frac{\ell^2}{r^2} \left(1-\frac{2M}{r}\right),
\end{align}
which, in the case $\ell=0$ pertaining to radially infalling geodesics, implies that   the photon four-velocity becomes 
\begin{align}
k^\mu = \Omega\left[\frac{1}{\F \varepsilon \left(1-2M/r\right)},  \frac{-1}{\sqrt{\varepsilon} \sqrt{\F^2}},0,0 \right],
\label{photon-four-velocity}
\end{align}
 the temporal component $k^t$  being computed by means of Eq. \eqref{Omega-photon}. 

We can now obtain a constraint on the $\F $ function by considering the photon radial velocity $\dd r / \dd t$ measured by an observer stationed at infinity. This quantity reads as 
\begin{align}
\frac{\dd r}{ \dd t }=\frac{k^r}{k^t}= \frac{\dd r}{\dd \lambda} \left(\frac{\dd t }{\dd \lambda}\right)^{-1} = - \sqrt{\varepsilon} \left(1-\frac{2M}{r}\right) \frac{\F }{\sqrt{\F ^2}},
\label{dr-dt-photon-1}
\end{align}
when Eq. \eqref{photon-four-velocity} is used, while it is given by 
\begin{align}
\frac{\dd r}{\dd t} = - \sqrt{\varepsilon} \left(1-\frac{2M}{r}\right),
\label{dr-dt-photon-2}
\end{align}
if the condition $\dd s^2 =0$ is exploited (see Eq. \eqref{Lorentzian-Euclidean-Schwarzschild}). Therefore, a comparison between formulas \eqref{dr-dt-photon-1} and \eqref{dr-dt-photon-2} yields 
\begin{align}
\F= \sqrt{\F^2}. 
\label{F-function-equation}    
\end{align}
As we will show in Sec. \ref{Sec:F-function}, the validity of the above equation can be confirmed by employing the signature-adaptive Lorentzian-Euclidean transformations developed in Sec. \ref{Sec:generalized-transformation}. 

The relation \eqref{F-function-equation} permits to  inspect the behavior of photons radially approaching the black hole. Similarly to the massive-particle scenario (cf. Eq. \eqref{E-alpha-varepsilon-2}),  from Eq. \eqref{Omega-photon} we see that the photon energy can be written as
\begin{align}
\Omega = \alpha_{\rm p} \, \F \varepsilon,
\label{Omega-expression}
\end{align}
for some positive-definite bounded function $\alpha_{\rm p}$ (notice that we are implicitly assuming $\F$ to be positive for $r>2M$, ensuring  that $\Omega$ is also positive in this region; we will come back on this point in Sec. \ref{Sec:F-function}). Therefore, it follows from Eq.  \eqref{photon-four-velocity}, that
\begin{align}
k^r= - \frac{\Omega}{\sqrt{\varepsilon} \sqrt{\F^2}}= - \alpha_{\rm p} \frac{\varepsilon}{\sqrt{\varepsilon}} \frac{\F}{\sqrt{\F^2}}= - \alpha_{\rm p} \sqrt{\varepsilon},
\label{k-r-photon-1}
\end{align}
where we have exploited Eq. \eqref{F-function-equation}. Bearing in mind Eq. \eqref{epsilon-of-r}, we thus find that  the behavior of photons mirrors that  of massive particles, as   $k^r$ vanishes on the event horizon and attains  imaginary values inside it.  The same conclusion can also be drawn  by examining the coordinate-time radial  velocity $\dd r / \dd t$, which can be read off from Eq. \eqref{dr-dt-photon-2}. This implies that  from the standpoint of both a distant observer and the photon itself, the singularity at $r=0$ is never reached by following null paths.  Consequently,  the concept of atemporality, which in our picture is viewed as the mechanism that regularizes the black hole geometry and allows one to avoid the singularity, can be extended to include  null geodesics as well. This means that  neither massive nor massless geodesics, i.e. causal geodesics,  can cross the event horizon.

Integrating Eq. \eqref{k-r-photon-1} for any finite initial position $r_0 >2M$, we find
\begin{align}
r(\lambda)= - \alpha_{\rm p} \sqrt{\varepsilon} \lambda + r_0,
\end{align}
where we have assumed that $\int  \dd \lambda \alpha_{\rm p} \sqrt{\varepsilon} =  \alpha_{\rm p} \sqrt{\varepsilon} \int  \dd \lambda $.  Therefore, photons get to the event horizon at the affine parameter value $\lambda^\star$
\begin{align}
\lambda^\star= \frac{r_0-2M}{\alpha_{\rm p} \sqrt{\varepsilon}} \xrightarrow[\varepsilon \to 0^+]{} +\infty.    
\label{lambda-star}
\end{align}
This result shows that null geodesics exhibit the same behavior as  timelike ones, as the event horizon is reached only in the limit of infinite affine parameter. This permits concluding that  the Lorentzian-Euclidean Schwarzschild spacetime is geodesically complete and the central singularity can be avoided. In Ref. \cite{Caponio2025}, an alternative discussion of Lorentzian-Euclidean Schwarzschild spacetime is reported.

A final comment is now in  order. At first glance, the existence of  photons with zero velocity might appear nonphysical. However, this phenomenon arises  as a  direct consequence of the degenerate character of the metric. Let us describe this point by considering  a photon  moving, for simplicity, in the $x$-direction of  flat Minkowski spacetime, with four-velocity $k^a = \left(k^0,k^x,0,0\right)$. In conventional Special Relativity, a photon never interrupts its motion   otherwise the relation $k^a k_a=0$ would not be satisfied. On the other hand, the presence, in our framework, of the flat degenerate metric $f_{ab}$ (see Eq.  \eqref{flat-metric}) implies 
\begin{align}
0= f_{ab} k^a k^b=   k^a k_a  = -\varepsilon \left(k^0\right)^2 + \left(k^x\right)^2, 
\label{photon-flat-case}
\end{align}
which yields $k^x=0$ when $\varepsilon=0$. When considering the Lorentzian-Euclidean geometry, the photon velocity becomes (asymptotically) zero only for $r=2M$, as we have just shown. In both the flat metric case and the Lorentzian-Euclidean black hole setting, this outcome can be understood in terms of the Lorentzian-Euclidean transformations discussed before, which are, by construction,  ill-defined when $\varepsilon=0$. Therefore, having photons with zero velocity makes complete sense in our model.

\subsection{The $\F $ function} \label{Sec:F-function}

In this section, we further examine the role of $\F $ function occurring in Eq. \eqref{Omega-photon} by exploiting the unified Lorentzian-Euclidean transformations devised earlier. For this reason, we first calculate the frequency $\omega^{(\rm s)}$ of the radially moving photon as determined by the stationary observer in her/his proper reference frame $\{\boldsymbol{e}_{a}\}$. Following the tetrad formalism outlined in Sec. \ref{Sec:Local-Lorentzian-Euclidean-transf},   we can write
\begin{align}
 \omega^{(\rm s)} =-\F \, \boldsymbol{k} \cdot \boldsymbol{e}_{\hat t}= -\F g_{\mu \nu} k^\mu u^\nu_{\rm s} = \frac{\Omega}{\sqrt{1-2M/r}},   
 \label{omega-s}
\end{align}
where we have taken into account Eqs.  \eqref{static-oberver-velocity}, \eqref{e-hat-t},  and \eqref{photon-four-velocity}. This equation has a crucial significance, as it proves that $\Omega$ can be interpreted similarly to the massive case, i.e. as the total energy of a  photon   relative to the static observer at infinity.

We can now compute the photon frequency $\omega^{(\rm f)}$ as evaluated by the geodetic observer. Referring to  the frame $\{\boldsymbol{e}_{a^\prime} \}$ carried by this observer, $\omega^{(\rm f)}$  is given by the projection of $\boldsymbol{k}$ on the timelike basis four-vector $\boldsymbol{e}_{\hat{0}^\prime} = \boldsymbol{u}$ (cf. Eq. \eqref{e-hat-zero-prime}), thus  yielding 
\begin{align}
\omega^{(\rm f)}&= -\F \, \boldsymbol{k} \cdot \boldsymbol{e}_{\hat{0}^\prime}= - \F g_{\mu \nu} k^\mu u^\nu
\nonumber \\  
&= \frac{\Omega}{\left(1-2M/r\right)} \left(\frac{E}{\varepsilon^2} - \frac{\F }{\sqrt{\F^2}} \sqrt{\frac{2M}{r}-1+\frac{E^2}{\varepsilon^4}}\right),
\label{omega-f-1}
\end{align}
where we have used Eq. \eqref{free-fall-velocity}. 

The measurements of $\omega^{(\rm s)}$ and $\omega^{(\rm f)}$ can be related via the generalized spacetime transformations worked out in Sec. \ref{Sec:generalized-transformation}. In fact, in view of Eq. \eqref{gamma-factor} and the structure of the local frame  change \eqref{general-transformations-curved}, $\omega^{(\rm s)}$ and $\omega^{(\rm f)}$ can be connected by the relation
\begin{align}
\omega^{(\rm f)}  = \omega^{(\rm s)} \sqrt{\frac{1+\mathscr{V}/\sqrt{\varepsilon}}{1-\mathscr{V}/\sqrt{\varepsilon}}} = \omega^{(\rm s)} \, \gamma \left(1 + \frac{\mathscr{V}}{\sqrt{\varepsilon}}\right),
\end{align}
which resembles the ordinary Doppler shift of gravity theory (see e.g. Ref. \cite{Mazzitelli2020} for the standard analysis framed in General Relativity). By means of Eqs. \eqref{boost-velocity-explicit} and \eqref{gamma-explicit}, we find
\begin{align}
\omega^{(\rm f)} = \frac{\Omega}{\left(1-2M/r\right)} \left(\frac{E}{\varepsilon^2} - \sqrt{\frac{2M}{r}-1+\frac{E^2}{\varepsilon^4}}\right),
\label{omega-f-2}
\end{align}
and hence a comparison with Eq. \eqref{omega-f-1} gives $ \F /\sqrt{\F^2}=1$, in agreement with Eq. \eqref{F-function-equation}. This  is  a key result, as it  confirms the effectiveness of the  Lorentzian-Euclidean transformations. 

Upon exploiting Eq. \eqref{F-function-equation}, it is now straightforward to see that the frequency $\tilde{\omega}^{(\rm f)}$, measured by the free-fall observer whose motion  begins  at rest from infinity, is given by   (see Eq. \eqref{free-fall-velocity-infinity}) 
\begin{align}
    \tilde{\omega}^{(\rm f)}&= -\F \, \boldsymbol{k} \cdot \tilde{\boldsymbol{u}}= - \F g_{\mu \nu} k^\mu \tilde{u}^\nu= \frac{\Omega}{1+\sqrt{2M/r}},
\end{align}
which can also be derived from Eq. \eqref{omega-f-2} by setting $E=\varepsilon^2$ (recall in fact that this represents the rest energy  of the geodetic particle).

Having double-checked the relation  \eqref{F-function-equation}, we can now address  its resolution. Driven by the symmetries of the black hole metric \eqref{Lorentzian-Euclidean-Schwarzschild} and the form of Eq. \eqref{epsilon-of-r}, a nontrivial solution can be represented by 
\begin{align}
    \F = \sqrt{\varepsilon} = 
\left \{ \begin{array}{rl}
& 1, \qquad \;\; \, {\rm if}\; r>2M,\\
& 0,  \qquad \;\; \, {\rm if}\;  r=2M,\\
&   \; \ii, \qquad \;\; \, {\rm if}\;  r<2M,\\
\end{array}
\right. \; .
\label{F-solution-1}
\end{align}
Therefore, differently from the dynamics of massive particles, where the energy features the real-valued factor $\varepsilon$, in the photon case, there exists a class of solutions of Eq. \eqref{F-function-equation} that can be imaginary in the Euclidean domain $\mathscr{D}_-$, as the above equation shows. Notice that Eq. \eqref{F-solution-1} satisfies the essential requirement of yielding the correct sign for $\Omega$  at least in the Lorentzian regime $\mathscr{D}_+$ (cf. Eq. \eqref{Omega-photon}).    

Relation \eqref{omega-s} represents the counterpart of  ordinary blueshift formula holding in General Relativity \cite{Mazzitelli2020}. Like in the standard case, $\omega^{\rm (s)}$ diverges at the event horizon, since, upon using Eq. \eqref{Omega-expression}, we have
\begin{align}
\lim_{r\to \left(2M\right)^+} \omega^{(\rm s)} = \lim_{r\to \left(2M\right)^+} \frac{\alpha_{\rm p} \, \F \varepsilon}{\sqrt{1-2M/r}} = + \infty. 
\end{align}
This conclusion applies independently of the  explicit form assumed by the solution \eqref{F-solution-1}, as the necessity for  $\F$ to take a finite positive value when $r>2M$ is the minimal condition it should comply with. However, by employing the regularized expression  \eqref{sign-regularization} for $\varepsilon$ jointly with Eq. \eqref{F-solution-1}, we again find that $\omega^{\rm (s)}$ blows up at the event horizon.  On the other hand, by applying an analogous reasoning, we recover the standard General Relativity result that $\tilde{\omega}^{\rm (f)}$   remains finite at the event horizon. The fact that the behavior of both $\omega^{(\rm s)}$ and $\tilde{\omega}^{\rm (f)}$ mirrors that predicted by  Einstein gravity  once again indicates that our model is able to reproduce the  conventional features characterizing the region outside the Schwarzschild black hole. 

Finally, it should be stressed that all findings  in this section rely solely on the constraint \eqref{F-function-equation} and not on the actual solution \eqref{F-solution-1}, thereby aligning with the  analysis of radial null geodesics from the previous section. 

\section{Causal structure} \label{Sec:Penrose-diagrams}

Having analyzed radial null geodesics, an interesting topic for further discussion regards Penrose diagrams, which are a precious tool for understanding the causal structure of a spacetime  (see e.g. Refs. \cite{Hawking1973-book,Wald-book1984,Townsend1997,Strominger1994,Carroll2004,Poisson2009,Blau}). We begin  our investigation with the Lorentzian-Euclidean flat spacetime (see Sec. \ref{Sec:Penrose-flat}), which sets the stage for the analysis of the Lorentzian-Euclidean black hole in Sec. \ref{Sec:Penrose-black-hole}. 

\subsection{Flat spacetime} \label{Sec:Penrose-flat}

The Lorentzian-Euclidean flat spacetime has been introduced in Sec. \ref{Sec:Lorentz-and-rotation}. Its signature-changing metric \eqref{flat-metric}  can be expressed  in polar coordinates as
\begin{align}
\dd s^2_{\rm flat}= - \varepsilon \dd t^2 + \dd r^2 + r^2 \dd \Omega^2.
\label{flat-metric-2}
\end{align}
Let us introduce some finite radial distance $r_0>0$ and assume  that $\varepsilon$ takes values $\varepsilon=1,0,-1$ for $r>r_0$, $r=r_0$, and $r<r_0$, respectively. 

In the standard Minkowski spacetime $\mathbb{R}^{1,3}$,  light cones are determined by the relation $\dd s_{\rm M}^2 =-\dd t^2 + \dd r^2 + r^2 \dd \Omega^2 =0$, which leads to the familiar double-cone structure in 1+3 dimensions. On the other hand, in the Euclidean space $\mathbb{R}^4$, a \qm{lightlike} condition analogous to the Minkowski case would be given by  $\dd s_{\rm E}^2 =\dd t^2 + \dd r^2 + r^2 \dd \Omega^2  =0$, which yields no real solutions for nontrivial displacements. This implies that $\mathbb{R}^4$ lacks real null directions and light cones cannot be defined. Consequently, the causal structure characterizing Minkowski spacetime has no direct counterpart in Euclidean space, and hence the Penrose diagram for the Lorentzian-Euclidean flat geometry \eqref{flat-metric-2} should exclude (at least) the region $r<r_0$.

As anticipated before (cf. Eq. \eqref{photon-flat-case}),  photons evolving in the Lorentzian-Euclidean flat spacetime exhibit a vanishing velocity when $\varepsilon=0$, i.e., at $r=r_0$. In spherical coordinates, the slope of light cones is represented by $\dd t / \dd r = \pm 1/\sqrt{\varepsilon}$. Therefore, as long as $r>r_0$, radial light cones have an inclination of  $\pm 45^{\circ}$, while for $r<r_0$ we recover the result   that null rays  follow  imaginary paths. However, when $\varepsilon=0$,  $\dd t / \dd r $ diverges,  meaning that the trajectory of radial photons coincides with the vertical line $r=r_0$. This differs from the conventional Minkowski framework, where vertical lines always correspond to the timelike geodesics of static observers. 

In the Lorentzian-Euclidean scenario, light cones  fold up at $r=r_0$, meaning that outgoing rays no longer move in the direction of increasing $r$. Any  signal emitted at $r=r_0$ remains trapped there indefinitely. This situation is reminiscent of what occurs in standard Schwarzschild geometry at the event horizon. However, in that case, the pathological behavior of light cones at $r=2M$ arises because Schwarzschild coordinates are not suitable for describing the physics at the Schwarzschild radius. By contrast, in the Lorentzian-Euclidean setup \eqref{flat-metric-2},  the structure of light cones at $r=r_0$ reflects the degenerate character of the metric and cannot be cured by resorting to a different set of coordinates.  We can thus conclude that a Penrose diagram is not meaningful at $r=r_0$, as the fundamental requirement that null rays be represented by $45^{\circ}$ lines does not hold.  

Given the above premises, we can thus satisfactorily draw a Penrose diagram only if $r>r_0>0$. Thus, following the usual procedure, we first introduce the null  coordinates 
\begin{align}
\tilde{u} =t-r, \qquad \tilde{v}=t+r,
\end{align}
with ranges 
\begin{align}
-\infty< \tilde{u} < \tilde{v} <+\infty,    \label{range1}
\end{align}
and then define the light cone variables
\begin{align}
\tilde{U}= \arctan \tilde{u}, \qquad    \tilde{V}= \arctan \tilde{v},
\end{align}
which are then subject to the triangle bound
\begin{align}
    -\pi/2 < \tilde{U} < \tilde{V} < \pi/2. \label{range2}
\end{align}
Thus, in the domain $r>r_0$,    the metric \eqref{flat-metric-2} is found to be conformally equivalent to
\begin{align}
\widetilde {\dd s}^2_{\rm flat} = - \dd T^2 + \dd R^2 + \sin^2 R \, \dd \Omega^2,
\label{conformal-metric-flat}
\end{align}
where $T$ and $R$ are given by
\begin{align}
T=\tilde{U} + \tilde{V}, \qquad R=\tilde{V}-\tilde{U},
\end{align}
and satisfy the constraints
\begin{subequations}
\label{range3}
\begin{align}
& \vert T \vert + R < \pi, \label{range3-a}
\\
& 0<R_0 < R < \pi.  \label{range3-b}
\end{align}  
\end{subequations}

In  the ordinary Minkowski setup, Eqs. \eqref{range1}, \eqref{range2}, and \eqref{range3-b} are slightly modified to include  $r=0$, and take the form:   $-\infty<\tilde{u} \leq  \tilde{v}<+\infty$, $ -\pi/2<\tilde{U} \leq \tilde{V}<\pi/2 $,  $0 \leq R < \pi$. In Eq. \eqref{range3-b},  $R_0$ is a parameter defined in such a way that if  $R>R_0$, then the lower bound $r>r_0$ is satisfied. 

The relation between $R_0$ and $r_0$ can be derived as follows. First, it easily follows from  the above formulas that the line $r=r_0$ becomes in the $(R,T)$-plane the curve
\begin{align}
\vert T \vert = \arccos \left(\frac{\sin R}{r_0}-\cos R\right).
\label{r-const-curve}
\end{align}
In general,  this curve  intersects  the vertical line $R=R_0$ at  one,   two,  or zero points. The first situation can only occur when $ T =0$. We thus find that this unique intersection point lies on the $R$-axis at
\begin{align}
R_0 = 2 \arctan r_0, \label{R0-r0-relation}
\end{align}
where, in our hypotheses, $0<R_0<\pi$ (cf. Eq. \eqref{range3-b}) as we have assumed that $r_0$ represents a finite radial distance.

\begin{figure}[bht!]
\centering\includegraphics[scale=0.30]{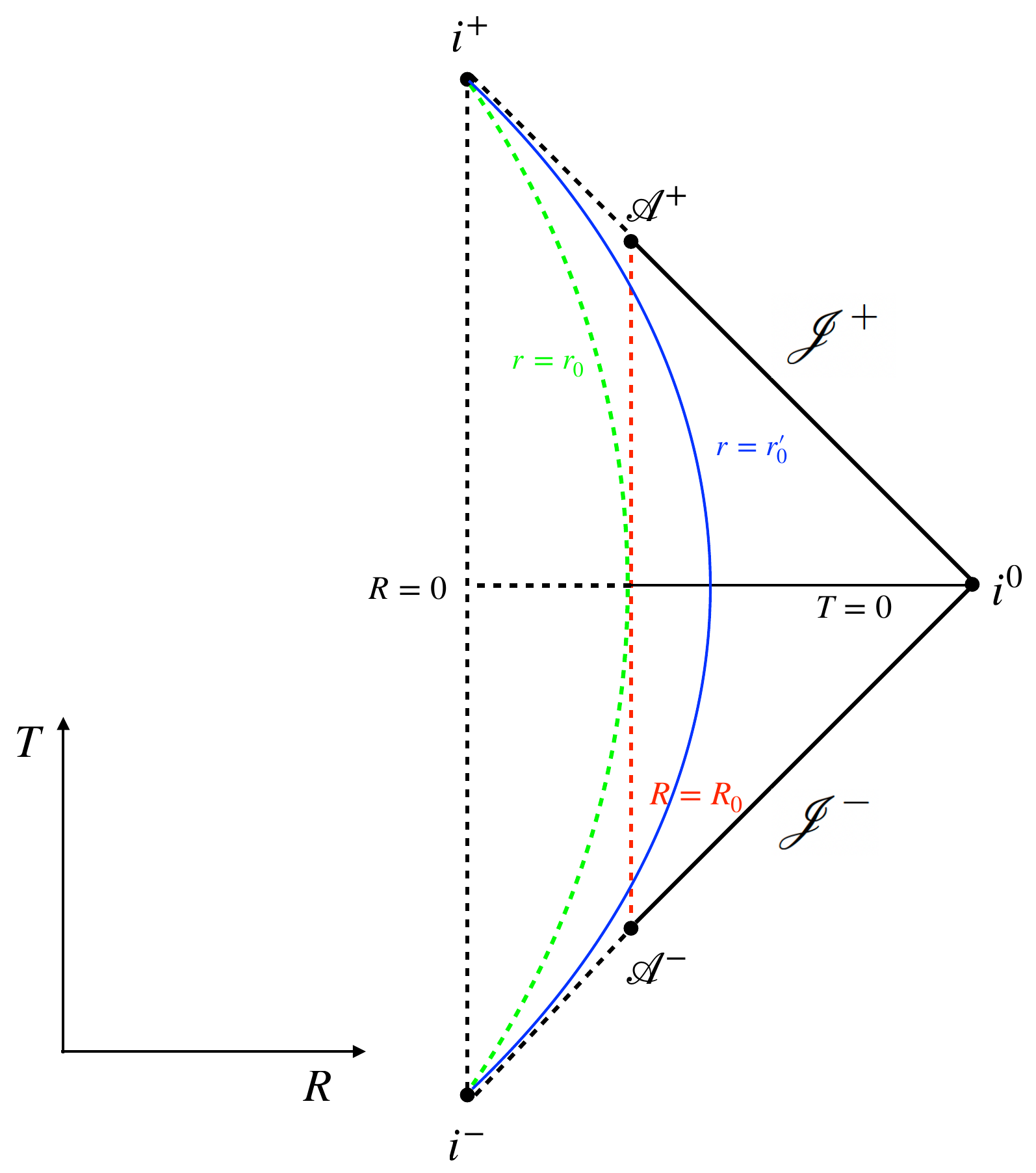}\hspace{1.25cm}
\caption{The Penrose diagram for the Lorentzian-Euclidean flat geometry \eqref{flat-metric-2} satisfying  the conditions \eqref{range3} and \eqref{R0-r0-relation}. The degenerate character of the metric implies that $i^{\pm}$ and a portion of $\mathscr{J}^{\pm}$ do not belong to the diagram. }
\label{Fig-Penrose-flat}
\end{figure}

We can thus conclude that the condition $r>r_0$ is fully satisfied when we impose the bound $R>R_0$, provided that  $R_0$ is given by Eq. \eqref{R0-r0-relation} and the range of $T$ is adjusted accordingly by means of Eq. \eqref{range3-a}. The Penrose  diagram for the Lorentzian-Euclidean flat geometry \eqref{flat-metric-2} complying with these requirements is shown in Fig. \ref{Fig-Penrose-flat}. Here, as usual, we define the future and past timelike infinities, $i^{+}$ and $i^-$, respectively, along with the spatial infinity,  $i^0$, and the future and past null infinities, $\mathscr{J}^+$ and $\mathscr{J}^-$, respectively, as follows:
\begin{subequations}
\begin{align}
 & \; \; \;\, i^+ \, : \, \left(T=\pi,R=0\right), 
\\
& \; \; \;\;\, i^0 \, : \, \left(T=0,R=\pi\right), 
\\
&\; \; \;\, i^- \, : \, \left(T=-\pi,R=0\right), 
\\
& \mathscr{J}^+ \, : \, \left(T=-R+\pi,0<R<\pi \right),
\\
& \mathscr{J}^- \, : \, \left(T=R-\pi,0<R<\pi \right).
\end{align}    
\end{subequations}

An interesting  criterion for identifying  a degenerate metric, at least in the flat case, is immediately evident, as $i^{\pm}$, along with  a portion of $\mathscr{J}^{\pm}$, do not  belong to the diagram. This follows from the fact that, in Minkowski spacetime, $i^{\pm}$   contain the future and past endpoints of \emph{all} (infinitely extended) timelike geodesics, while $\mathscr{J}^{\pm }$ represent the future and past endpoints of \emph{all} (infinitely extended) outgoing and ingoing radial null geodesics. On the other hand,  in the Lorentzian-Euclidean framework, this structure breaks down due to the absence of light cones and the underlying timelike (in the Lorentzian sense) trajectories in the Euclidean zone $r<r_0$. 

Moreover, as pointed out before, the curve $r=r_0$ corresponds  to the trajectory of a photon. Thus, if $i^{\pm }$ were included in the Penrose diagram, this  would imply the existence of a null geodesic evolving between $i^-$ and $i^+$, a result that does not align with the standard interpretation of these points. However, if we allow, for a moment, a potential generalization of the usual Penrose diagram,  the  metric degeneracy could also be understood as a  spacetime  property that   entails the presence of (at least) one null geodesic running between  $i^-$ and  $i^+$.

All admissible timelike geodesics cross the line $R=R_0$ twice. This is  illustrated in Fig. \ref{Fig-Penrose-flat} by  the curve $r=r_0^\prime$,  with $r_0^\prime$ a constant larger than  $r_0$. The \qm{time} coordinates of the intersection points always lie between the limiting values $R_0-\pi$  and $-R_0 + \pi$, which correspond to the   \qm{time} positions of  $\mathscr{A}^{-}$ and $\mathscr{A}^{+}$, respectively. This scenario suggests,  in some  sense, that we can think of    $i^{\pm}$ as being replaced by  $\mathscr{A}^{\pm} \equiv \left(R=R_0, T=\mp R_0 \pm \pi\right)$. However, it is important to recall  that the former are really points, while the latter are two-spheres, as the \qm{radius} $\sin R_0$  does not vanish at $\mathscr{A}^{\pm}$ (cf. Eq. \eqref{conformal-metric-flat}).

In the ordinary  framework, any timelike observer has access to the whole Minkowski spacetime, as the past and future light cones of a  timelike observer located at $i^{\pm}$  cover the entire  spacetime. This does not hold  in our model, where the timelike observer is able to see only  the Lorentzian domain of the Lorentzian-Euclidean spacetime. This situation, which interestingly bears a resemblance to black hole geometries, can be easily explained through the aforementioned \qm{points} $\mathscr{A}^{\pm}$, as only the Lorentzian portion of the  Lorentzian-Euclidean flat space is in the causal past (resp. future) of an observer located  at  $\mathscr{A}^{+}$ (resp.  $\mathscr{A}^{-}$).   Finally, the breakdown of causal propagation for $r<r_0$  implies that the Lorentzian and Euclidean regions are not in causal contact.  This feature shares some similarities with  the Robertson-Walker universe with scale factor $a(t) \, \propto \, t^q$ for $0<q<1$ (in a sense, the Penrose diagram of Fig. \ref{Fig-Penrose-flat} can be viewed as the \qm{reversed} version  of the Robertson-Walker spacetime diagram, with our case being valid for $r>r_0$, while the cosmological one for  $t>0$), and will also appear  in the curved case investigated in the next section.

\subsection{Black hole spacetime} \label{Sec:Penrose-black-hole}

In this section, we examine how the presence of the degenerate surface  $r=2M$ affects the global causal structure of the Lorentzian-Euclidean Schwarzschild geometry. We first work out the Kruskal extension of the metric in Sec. \ref{Sec:Kruskal}\footnote{We recall that  Kruskal-Szekeres coordinates were derived in the Appendix A of Ref. \cite{Capozziello2024} under the simplified assumption $\varepsilon=\pm 1$;  in this section, however, we take into account the distributional nature of $\varepsilon$.}, and then construct the Finkelstein and Penrose diagrams in  Secs. \ref{Sec:Finkelstein-diagrams} and \ref{Sec:Causal-structure}, respectively. Finally, we discuss the properties of the event horizon in  Sec. \ref{Sec:properties-event-horizon}.

\subsubsection{Kruskal extension}\label{Sec:Kruskal}

Inspired by the standard approach (see e.g. Refs. \cite{MTW,Hawking1973-book,Chandrasekhar1985,Carroll2004,Poisson2009,Blau}), we express the metric \eqref{Lorentzian-Euclidean-Schwarzschild} as
\begin{align}
    \dd s^2 = \left(1-\frac{2M}{r}\right) \left[-\varepsilon \dd t^2 + \frac{\dd r^2}{\left(1-2M/r\right)^2}\right] + r^2 \dd \Omega^2,
\end{align}
which suggests  defining the tortoise coordinate $r^\star$ in the usual way: 
\begin{align}
r^\star= \int \frac{\dd r}{\left(1-2M/r\right)} = r + 2M \log \left(\frac{r}{2M}-1\right),
\label{r-star}
\end{align}
thus yielding
\begin{align}
\dd s^2 = \left(1-\frac{2M}{r}\right) \left(-\varepsilon \dd t^2 + \dd r^{\star 2}\right) + r^2 \dd \Omega^2.  
\label{metric-r-star}
\end{align}
In the new coordinate system, radial null geodesics satisfy
\begin{align}
\sqrt{\varepsilon} \dd t = \pm \dd r^\star,
\label{dt-dr-star}
\end{align}
and hence follow real paths for $\varepsilon=1$ and  imaginary ones if $\varepsilon=-1$. Moreover, unlike the ordinary Lorentzian solution, the quantity $\dd t/ \dd r^\star  $ becomes pathological at the event horizon due to the degeneracy of the metric, which is not a mere coordinate singularity. In other words, the tendency of light cones to close up at $r=2M$ cannot be simply removed via a coordinate transformation in our setup. It is worth noticing  that $r^\star $ is originally defined only for $r>2M$, where $\varepsilon=1$,  but it can be formally extended  to $r<2M$  by  taking the modulus in the logarithmic term of Eq. \eqref{r-star}, although this procedure yields  a multivalued function.  

Let us introduce a set of generalized null coordinates 
\begin{subequations}
\label{null-coord-u-v-1}
\begin{align}
\bar{u}&= \sqrt{\varepsilon} t - r^\star,
\label{null-coord-u-1} 
\\
\bar{v}&= \sqrt{\varepsilon} t + r^\star,
\label{null-coord-v-1} 
\end{align}    
\end{subequations}
then from Eq. \eqref{dt-dr-star} jointly with the identity
\begin{align}
\dd \left(t \sqrt{\varepsilon}\right)= \sqrt{\varepsilon} \dd t,     
\label{relation-diff}
\end{align}
we see that outgoing and ingoing light rays are described by $\bar{u}=const$  and $\bar{v}=const$, respectively. 

The validity of Eq.  \eqref{relation-diff} can be established in two ways. First, by the direct  computation
\begin{align}
\dd \left(t \sqrt{\varepsilon}  \right) = 
\left \{ \begin{array}{rl}
& \dd t, \qquad \;\; \, {\rm if}\; r>2M,\\
& 0,  \qquad \;\; \; \,\, {\rm if}\;  r=2M,\\
& \ii \dd t , \qquad \;\; {\rm if}\;  r<2M,\\
\end{array}
\right. \Rightarrow \dd \left(t \sqrt{\varepsilon}  \right)=\sqrt{\varepsilon} \dd t ,
\end{align}
and second, by pursuing  a more formal approach. Indeed,  relation  \eqref{relation-diff} implies that   $t \left(\dd \sqrt{\varepsilon}\right)$ gives a vanishing contribution in the distributional sense, a result that  can be demonstrated as follows. Proceeding along the same lines as in Ref. \cite{Capozziello2024},  our calculations  employ  the smooth approximation \eqref{sign-regularization} for terms involving $\varepsilon$, but not its derivatives, thereby allowing us to keep track of its distributional character. Therefore, bearing in mind that all quantities involving a Dirac $\delta$ function should be interpreted as distributions to be integrated over smooth functions, we have
\begin{align}
\int t \left(\dd \sqrt{\varepsilon} \right)=&  2M \int \dd r \frac{t}{r^2} \frac{\delta\left(1-2M/r\right)}{\sqrt{\varepsilon}}  
= 2M \int \dd r \frac{t}{r^2} 
\nonumber \\
& \times \frac{\delta\left(1-2M/r\right)}{\left(r-2M\right)^{\frac{1}{2(2 \kappa +1)}}} 
 \left[\left(r-2M\right)^2 + \varrho\right]^{\frac{1}{4(2 \kappa +1)}},
\end{align}
which can be shown to be zero via the Hadamard regularization, which relies on the prescription  $\delta(x) \vert x \vert^{-n} \equiv 0$,  with $n$  a positive integer (see Sec. 9.6 in Ref. \cite{Poisson-Will2014} and Sec. III in Ref. \cite{Capozziello2024} for further details). 

In terms of the null variables \eqref{null-coord-u-v-1}, the metric  \eqref{metric-r-star} becomes
\begin{align}
\dd s^2 =- \left(1-\frac{2M}{r}\right) \dd \bar{u} \dd \bar{v} + r^2 \dd \Omega^2,
    \label{metric-in-out}
\end{align}
which reveals  a subtle feature of the Lorentzian-Euclidean geometry. Since  $\bar{u}$ and $  \bar{v}$ are real in the Lorentzian domain $\mathscr{D}_+$ but become complex in the Euclidean region $\mathscr{D}_-$, then for $r<2M$ Eq. \eqref{metric-in-out} appears as a complexified metric expressed in a  complex basis and its (complexified) signature is Lorentzian. In other words, although the underlying real signature in $\mathscr{D}_-$ is ultrahyperbolic, the metric \eqref{metric-in-out} takes a Lorentzian structure when regarded as a bilinear form on the complexified tangent space (hence the terminology complexified Lorentzian signature; see Ref. \cite{Garnier2025} for further details about complexified metrics).

Let us further analyze this point with a simple example. Consider the following flat ultrahyperbolic  metric: 
\begin{align}
\dd s^{\prime 2} = -\left(\dd t_1\right)^2 - \left(\dd t_2\right)^2 + \left(\dd t_3\right)^2 + \left(\dd t_4\right)^2,  
\label{metric-example}
\end{align}
with $t_1,t_2,t_3,t_4 \in \mathbb{R}$. Now, similarly to Eq. \eqref{null-coord-u-v-1},  by means of the complex null coordinates
\begin{align}
\mathcal{U}=\ii t_1-t_2,  \qquad \mathcal{V}=\ii t_1+t_2,  
\end{align}
 we can  write
\begin{align}
\dd s^{\prime 2} = \dd \mathcal{U}  \dd \mathcal{V}  +  \left(\dd t_3\right)^2 + \left(\dd t_4\right)^2,
\end{align}
which  formally represents a complexified metric with (complexified) Lorentzian signature. On the other hand,  defining the real (but not null) variables
\begin{align}
\mathcal{U}^\prime=t_1-t_2,  \qquad \mathcal{V}^\prime=t_1+t_2,  
\end{align}
allows the metric  to assume the diagonal form 
\begin{align}
\dd s^{\prime 2} =-\frac{1}{2}\left( \dd \mathcal{U}^{\prime 2} + \dd \mathcal{V}^{\prime 2}\right)  +  \left(\dd t_3\right)^2 + \left(\dd t_4\right)^2,
\end{align}
which  clearly exhibits ultrahyperbolic signature. The correct way to recover the full null structure of the geometry \eqref{metric-example} is to introduce two  independent sets of real null coordinates  involving linear combinations of $t_{1,3}$ and $t_{2,4}$: 
\begin{align}
& \tilde{\mathcal{U}}=t_1-t_3,  \qquad \tilde{\mathcal{V}}=t_1+t_3,
\nonumber \\
& \tilde{\mathcal{U}}^\prime=t_2-t_4,  \qquad \tilde{\mathcal{V}}^\prime=t_2+t_4,
\end{align}
in which  the metric is expressed purely in off-diagonal form
\begin{align}
 \dd s^{\prime 2}   = - \dd \tilde{\mathcal{U}} \dd \tilde{\mathcal{V}} - \dd \tilde{\mathcal{U}}^\prime \dd \tilde{\mathcal{V}}^\prime,
\end{align}
and displays ultrahyperbolic signature. 

This analysis shows that complex null coordinates can \qm{obscure} the actual  signature of the metric. In particular, there is no unique prescription for defining null coordinates across a signature-changing surface, especially in the presence of complexification. The choice of a basis directly influences how the signature manifests: the metric can appear Lorentzian when expressed in complex coordinates, even though its real signature is ultrahyperbolic. Therefore, care must be taken in interpreting such coordinates and the associated causal structure.

In view of these premises, the most natural way to derive the Kruskal-Szekeres form of the Lorentzian-Euclidean Schwarzschild metric is to  resort to  analytic continuation techniques, which have been successfully  used in various research fields  (e.g., Euclidean quantum gravity \cite{GH1977,Hawking1979-path,Esposito1994}). In such  frameworks, the link between the Lorentzian-signature Schwarzschild metric expressed in Kruskal-Szekeres coordinates $(\bar{T},\bar{R},\theta,\phi)$
\begin{align}
\dd s_{\rm L}^2 =\frac{32 M^3}{r} \ee^{-r/2M} \left(- \dd \bar{T}^2 + \dd \bar{R}^2\right) + r^2 \dd \Omega^2,
\label{Lorentzian-KS-metric}
\end{align}
and its Euclidean counterpart
\begin{align}
\dd s_{\rm E}^2 =\frac{32 M^3}{r} \ee^{-r/2M} \left( \dd \bar{T}^2 + \dd \bar{R}^2\right) + r^2 \dd \Omega^2,
\label{Euclidean-KS-metric}
\end{align}
which is formally valid for $r  \geq 2M$ \cite{GH1977},  is  established through the analytic continuation $\bar{T} \to \ii \bar{T}$, while the substitution $t \to \ii t$ is invoked when working in Schwarzschild coordinates.

We  thus propose a similar, though slightly  different, construction suitable for the Lorentzian-Euclidean Schwarzschild geometry, where the metric is given by Eq. \eqref{Lorentzian-KS-metric} if $r>2M$, while it becomes ultrahyperbolic when $r<2M$ (cf. Eq. \eqref{ultrahyperbolic-KS-metric} below). In our  approach, the interior geometry is derived  by performing the transformation
\begin{align}
\bar{T} \to \bar{T}, \qquad \bar{R} \to \ii \bar{R},  
\label{mapping-Kruskal}
\end{align}
on Eq. \eqref{Lorentzian-KS-metric} (in a sense, Eq. \eqref{mapping-Kruskal} appears as the mirrored version of the aforementioned standard mapping). The relations between Schwarzschild and Kruskal-Szekeres coordinates for $r>2M$ remain the standard Lorentzian expressions (regions I and III of the Kruskal diagram). For $r<2M$, however,  they follow from applying the  mapping \eqref{mapping-Kruskal}, jointly with the replacement $t \to \ii t$,  to the ordinary Lorentzian formulas pertaining  to   domains II and IV of the Kruskal diagram \cite{MTW}. In this way, borrowing the same nomenclature from the Lorentzian world,  we have   in the sectors I and III 
 \cite{MTW,Carroll2004}
\begin{align}
\underset{(r>2M)}{({\rm I}),  ({\rm III}):}
 \left \{  \begin{array}{rl}
& \bar{T}= \pm \ee^{r/4M} \sqrt{r/2M-1}  \,  \sinh \left(t/4M\right),  \\ [0.2cm]
& \bar{R}= \pm \ee^{r/4M} \sqrt{r/2M-1}  \,  \cosh \left(t/4M\right),  \\ [0.2cm]
\end{array}
 \right. 
 \label{I-III-relations}
\end{align}
while for the zones II and IV we get  the novel formulas 
\begin{align}
\underset{(r<2M)}{({\rm II}),({\rm IV}):} 
 \left \{  \begin{array}{rl}
& \bar{T}= \pm \ee^{r/4M} \sqrt{1-r/2M}  \,  \cos \left(t/4M\right),  \\ [0.2cm]
& \bar{R}= \pm \ee^{r/4M} \sqrt{1-r/2M}  \,  \sin \left(t/4M\right).  \\ [0.2cm]
\end{array}
 \right. 
 \label{II-IV-relations}
\end{align}
In the above equations,  the upper (resp. lower) sign refers to the domains I and II (resp.  III and IV), with  relations \eqref{I-III-relations}  valid if $r>2M$, while Eq. \eqref{II-IV-relations} for $r<2M$. One can verify that the transformations \eqref{I-III-relations} yield Eq. \eqref{Lorentzian-KS-metric}, while Eq. \eqref{II-IV-relations} leads to  the sought-after ultrahyperbolic metric 
\begin{align}
\dd s^2_{\rm II,IV} = \frac{32 M^3}{r} \ee^{-r/2M} \left(- \dd \bar{T}^2 - \dd \bar{R}^2\right) + r^2 \dd \Omega^2. 
\label{ultrahyperbolic-KS-metric}
\end{align}

It follows from Eq. \eqref{I-III-relations}, that lines of constant $r$ trace rectangular hyperbolas
\begin{align}
\bar{R}^2 - \bar{T}^2 = \left(\frac{r}{2M}-1\right) \ee^{r/2M},
\label{constant-r-hyperbolas}
\end{align}
while constant-$t$ surfaces are given by  straight lines 
\begin{align}
\frac{\bar{T}}{\bar{R}}= \tanh\left(\frac{t}{4M}\right). 
\label{constant-t-outside}
\end{align}
Conversely, inside the horizon, Eq. \eqref{II-IV-relations} shows that   constant-$r$ curves become  circles 
\begin{align}
\bar{R}^2 + \bar{T}^2 = \left(1-\frac{r}{2M}\right) \ee^{r/2M},
\label{constant-r-circles}
\end{align}
and 
\begin{align}
\frac{\bar{T}}{\bar{R}}= \cot\left(\frac{t}{4M}\right), 
\end{align}
describes isotemporal lines.

Although Eq. \eqref{I-III-relations} holds for $r>2M$, we see that the relation \eqref{constant-r-hyperbolas} suggests  identifying the event horizon with lines  $\bar{T}= \pm \bar{R}$ (i.e., the asymptotes of the rectangular hyperbolas), which intersect at the point $(\bar{R},\bar{T})=(0,0)$, where the right-hand side of  Eq. \eqref{constant-r-circles} vanishes. This choice implies that $t \to \pm \infty$ along the event horizon, as seen from  Eq. \eqref{constant-t-outside}. 

\begin{figure*}[bht!]
\centering\includegraphics[scale=0.40]{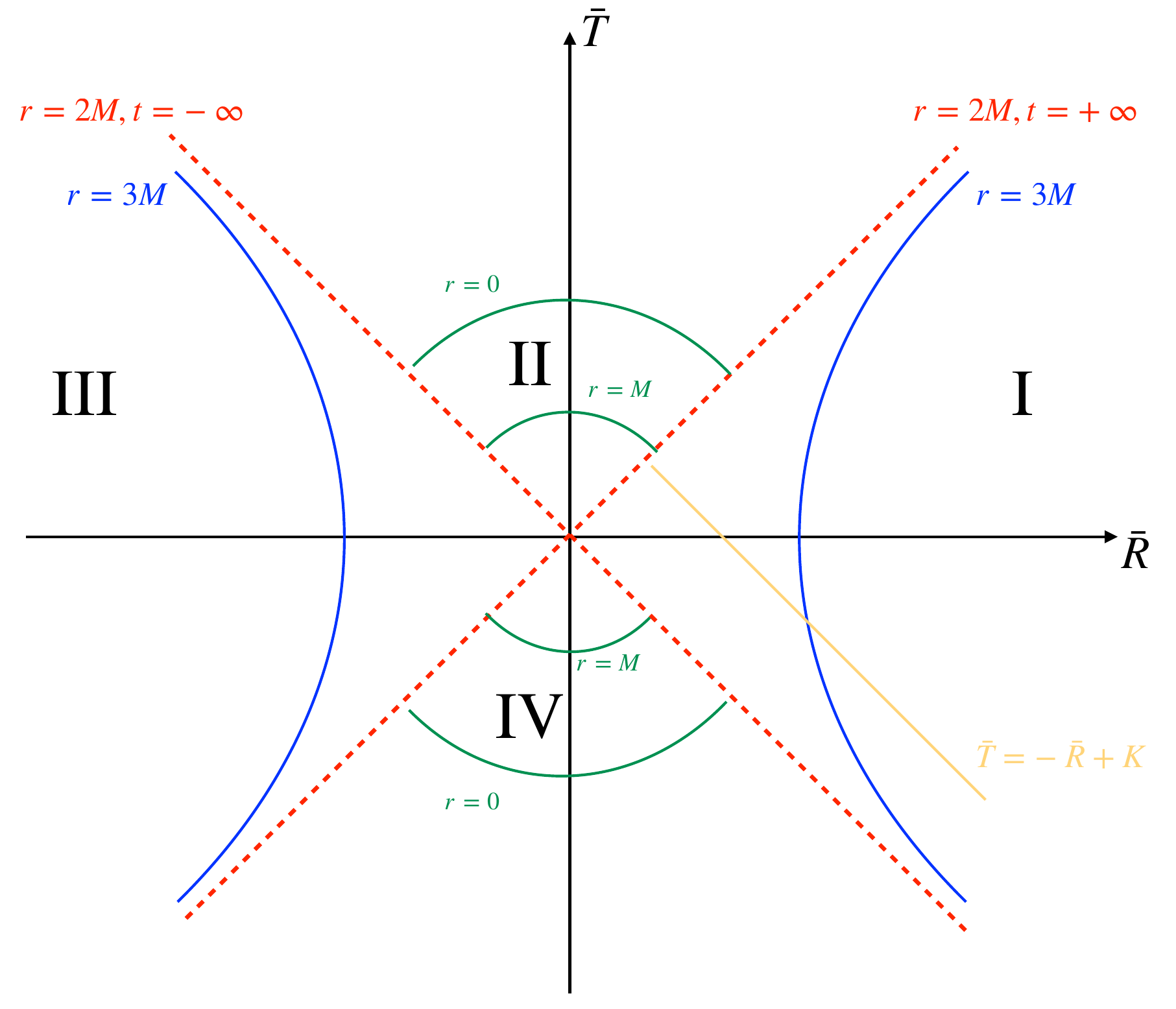}\hspace{1.25cm}
\caption{The Kruskal diagram of the Lorentzian-Euclidean Schwarzschild spacetime. In regions I and III,  the coordinates are described  by Eq. \eqref{I-III-relations} and the metric is given by Eq. \eqref{Lorentzian-KS-metric}, while in domains II and IV,  Eqs.  \eqref{II-IV-relations} and  \eqref{ultrahyperbolic-KS-metric} apply. Constant-$r$ curves are the hyperbolas \eqref{constant-r-hyperbolas} for $r>2M$ and  the circles \eqref{constant-r-circles} if $r<2M$.  The event horizon, where the metric becomes degenerate, corresponds to the straight lines $\bar{T}= \pm \bar{R}$, where $t \to \pm \infty$. An ingoing radial null geodesic  (yellow line $\bar{T}=-\bar{R} + K$, with $K$ a constant)  approaches $r=2M$ asymptotically and thus never intersects the red dashed line. }
\label{Fig-Kruskal}
\end{figure*}

Since the Lorentzian-Euclidean geometry becomes degenerate at $r=2M$, the spacetime  naturally decomposes into two disjoint regions, described by the metrics \eqref{Lorentzian-KS-metric} and \eqref{ultrahyperbolic-KS-metric},  and separated by the event horizon. The ensuing  Kruskal diagram is given in Fig.  \ref{Fig-Kruskal}. 

The Kruskal-Szekeres coordinates cover the entire spacetime manifold, except for the event horizon, according to the following scheme:
\begin{subequations}
\begin{align}
({\rm I})&: -\bar{R}<\bar{T}<\bar{R}, \; \bar{R}>0, 
\\
({\rm II})&: \bar{T} > \vert \bar{R} \vert, \; \bar{T}^2 + \bar{R}^2 <1, 
\\
({\rm III})&: \bar{R}<\bar{T}<-\bar{R}, \;  \bar{R}<0, 
\\
({\rm IV})&: \bar{T} < - \vert \bar{R} \vert, \; \bar{T}^2 + \bar{R}^2 <1.
\end{align}
\end{subequations}

As a consequence  of Eqs. \eqref{Lorentzian-KS-metric} and \eqref{ultrahyperbolic-KS-metric}, radial null geodesics  follow real paths only outside the event horizon. Specifically, we can track an ingoing light ray only while $r>2M$, which corresponds to the domain of validity of the outer Lorentzian metric \eqref{Lorentzian-KS-metric}. For this reason, the yellow line $\bar{T}=-\bar{R} + K$ (with $K$ a constant)  in Fig. \ref{Fig-Kruskal} does not touch the event horizon, as for $r=2M$ the geometry becomes degenerate, and the geodesic equations for radial photons are ill-defined. This indicates that the photon requires an infinite affine parameter to reach the event horizon (cf. Eq. \eqref{lambda-star}),  where it ultimately stops, as discussed in Sec. \ref{Sec:radial-photons}.

In addition, our model naturally predicts that no white hole can  form, as no physically admissible trajectory can emerge from the degenerate surface at $r=2M$ with nonzero velocity. This feature can be regarded as a positive outcome of the Lorentzian-Euclidean model, since in General Relativity white holes are often considered as nonphysical objects that cannot arise from collapsing matter \cite{Blau,Carroll2004}. As a consequence, the conventional notion of  past and future event horizons should be slightly modified. We will provide more details about this point in the next section, where we will deal with Finkelstein diagrams.

Prior to that, a final remark is in order. The Kruskal diagram displayed in Fig. \ref{Fig-Kruskal}  illustrates   the maximal analytic extension of the  Lorentzian-Euclidean spacetime. Unlike the corresponding Lorentzian-signature Schwarzschild solution, this geometry is geodesically complete, as any causal geodesic evolving in the Lorentzian domain reaches the event horizon at arbitrarily large  values of its affine parameter.

\subsubsection{Finkelstein diagrams}\label{Sec:Finkelstein-diagrams}

In light of the discussion of the last section regarding the nature of null coordinates, we can  provide  sensible Finkelstein diagrams for the Lorentzian-Euclidean geometry in the region $r \geq 2M$. Let us begin  by employing   ingoing Eddington-Finkelstein coordinates $(\bar{v},r,\theta,\phi)$, which allow  the metric \eqref{metric-r-star} to be expressed as
\begin{align}
    \dd s^2 = -\left(1-\frac{2M}{r}\right) \dd \bar{v}^2+ 2 \dd  \bar{v}  \dd r + r^2 \dd \Omega^2,
\end{align}
where we have made use of the identity \eqref{relation-diff}. It is  easy to see that an outward-directed light ray has four-velocity 
\begin{align}
  K^\mu_{\rm out} = \left(\frac{\dd {\bar{v}}}{\dd \lambda},\frac{\dd r}{\dd \lambda},\frac{\dd \theta}{\dd \lambda},\frac{\dd \phi}{\dd \lambda}\right) =\alpha_{\rm p} \sqrt{\varepsilon} \left(\frac{2}{1-2M/r},1,0,0\right),
\end{align}
which implies, on the one hand,  that  photons with positive radial velocity $\dd r/\dd \lambda$  exist as long as $r>2M$ (whereas for $r=2M$, we recover the earlier result that $\dd r/\dd \lambda=0$), and, on the other,    that $\dd \bar{v}/\dd r=\left(\dd \bar{v}/\dd \lambda\right) \left(\dd r / \dd \lambda\right)^{-1} $ blows up as $r \to 2M$, like in General Relativity.

\begin{figure}[bht!]
\centering\includegraphics[scale=0.25]{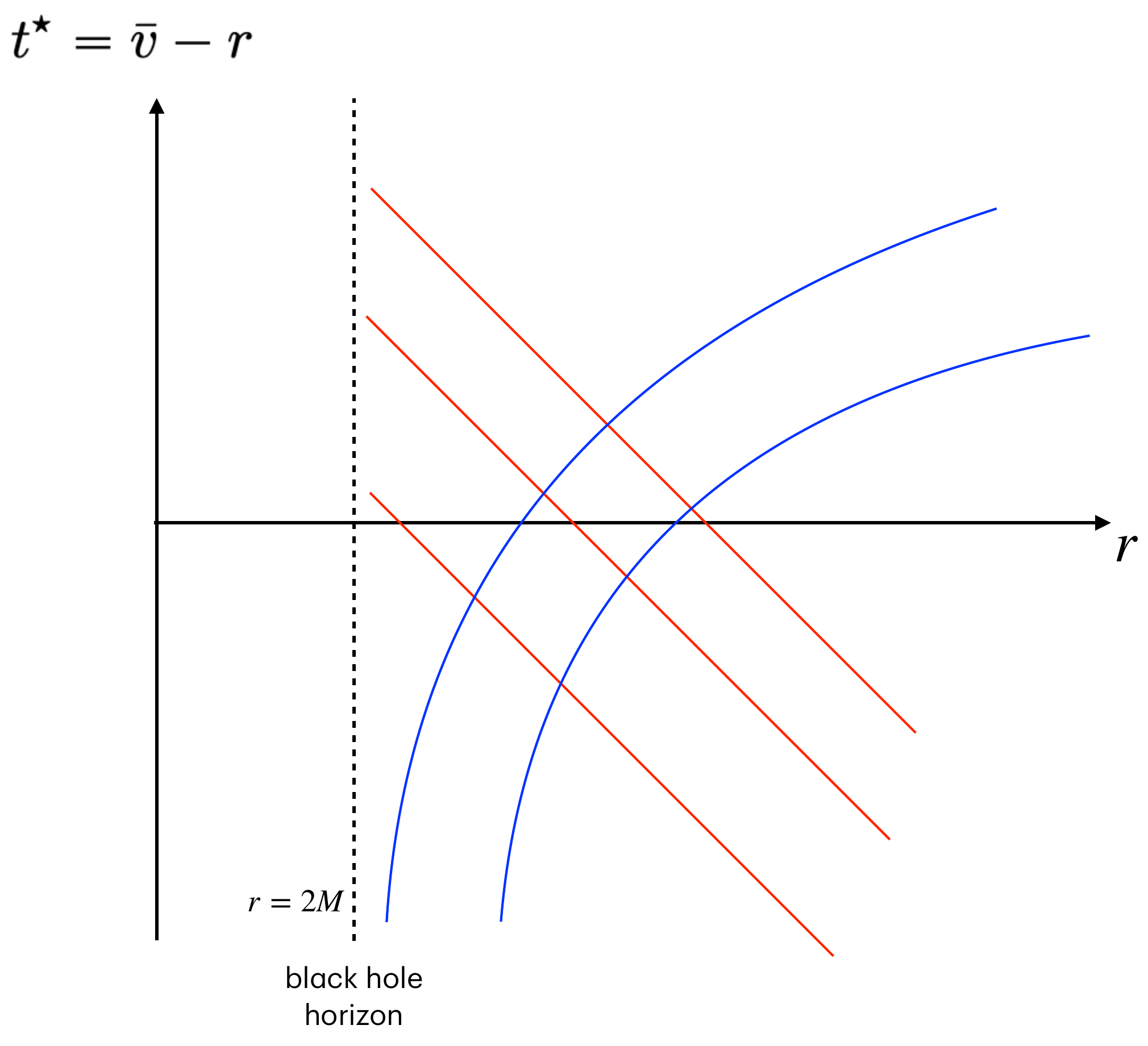}\hspace{1.25cm}
\caption{Finkelstein diagram in ingoing coordinates, with infalling radial null geodesics shown in red and outgoing ones in blue.  Both ingoing and outgoing light rays are unable to cross the (future) black hole event horizon. }
\label{Fig-Ingoing-Finkelstein}
\end{figure}

On the other hand, infalling null geodesics  are described, as long as $r>2M$,  by the relation $\bar{v}=\bar{v}_0$,  $\bar{v}_0$ being a constant.  At $r=2M$, however,  the definition \eqref{null-coord-v-1} entails  $\bar{v}=r^\star$, which in turn indicates that $\dd \bar{v}/\dd r$ again diverges on the event horizon. Therefore, both  inward- and outward-moving photons exhibit $\dd \bar{v}/\dd r \to \infty$ for $r=2M$, which points out their inability to cross the event horizon. While this behavior is well known in General Relativity for outgoing radial rays, its extension to ingoing ones is a distinctive feature of our model. The corresponding Finkelstein diagram is given in Fig. \ref{Fig-Ingoing-Finkelstein}, where we have defined $t^\star=\bar{v}-r$. 

Let us now consider outgoing Eddington-Finkelstein coordinates $(\bar{u},r,\theta,\phi)$, in which the metric takes the form
\begin{align}
    \dd s^2 = -\left(1-\frac{2M}{r}\right) \dd \bar{u}^2- 2 \dd  \bar{u}  \dd r + r^2 \dd \Omega^2.
\end{align}
In this case, infalling radial photons have four-velocity
\begin{align}
  K^\mu_{\rm in} = \left(\frac{\dd {\bar{u}}}{\dd \lambda},\frac{\dd r}{\dd \lambda},\frac{\dd \theta}{\dd \lambda},\frac{\dd \phi}{\dd \lambda}\right) =\alpha_{\rm p} \sqrt{\varepsilon} \left(\frac{2}{1-2M/r},-1,0,0\right),
\end{align}
while outgoing ones are characterized, when $r<2M$, by the condition $\bar{u}=\bar{u}_0$ (with $\bar{u}_0$ a constant), which boils down to $\bar{u}=-r^\star$ at $r=2M$ (see Eq. \eqref{null-coord-u-1}). 

\begin{figure}[bht!]
\centering\includegraphics[scale=0.25]{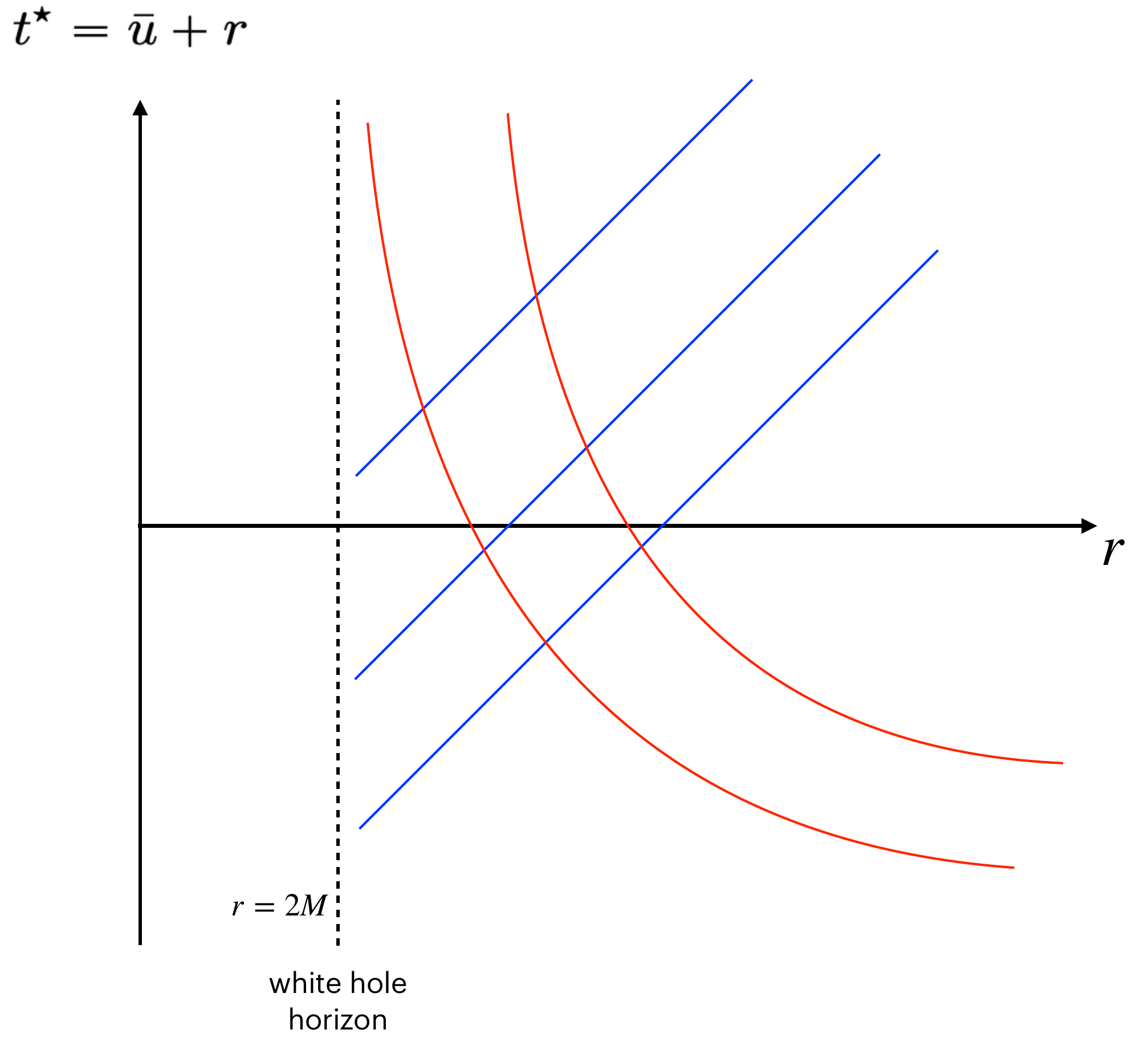}\hspace{1.25cm}
\caption{Finkelstein diagram in outgoing coordinates, with infalling radial null geodesics shown in red and outgoing ones in blue.  Both ingoing and outgoing light rays are unable to enter the (past) white hole event horizon. }
\label{Fig-Outgoing-Finkelstein}
\end{figure}

The  Finkelstein diagram is shown in Fig. \ref{Fig-Outgoing-Finkelstein}, where this time $t^\star=\bar{u}+r$. 
Like before, both outgoing and ingoing light rays exist for $r>2M$, yet neither is able to pass through the event horizon, where $\dd \bar{u}/\dd r$ diverges. This stands in stark contrast to Einstein gravity, where only infalling photons fail to traverse the event horizon and  appear to pile up at $r=2M$. 

Few comments should now be made. First of all, in line with the examination of Sec. \ref{Sec:Kruskal},  in the analysis performed in this section we have used expression \eqref{r-star} for the tortoise coordinate $r^\star$,  rather than its multivalued version involving the  modulus in the argument of the logarithm, which is nevertheless the conventional prescription   adopted in the Lorentzian framework for drawing Finkelstein diagrams. 

Secondly, our examination shows that the spacetime can be extended in two different directions, ensuring that the Lorentzian-Euclidean geometry features both (future) black hole and (past) white hole event horizons. However, since no trajectory enters the former and none emerges from the latter, their interpretation  differs from the ordinary one. In particular, the presence of the white hole, as typically conceived in Einstein theory,  is automatically ruled out. Remarkably, this absence arises not from an imposed boundary condition, but as a direct consequence of the degenerate geometry and the structure of null geodesics. This outcome could have been expected on general grounds. Indeed,  it stems from the fact that $\bar{u}$ and $\bar{v}$ represent sensible null coordinates adapted to null geodesics only in the domain $r>2M$. This fact entails   that if any signal were to originate from (or propagate into) the region $r<2M$, it would have to move along radial imaginary paths. Furthermore, if nothing crosses the future event horizon, as demonstrated before, then, by the time reversal invariance of the Lorentzian-Euclidean Schwarzschild solution, no information can come out of  the past horizon either. In other words, in our setup the future and past event horizons appear as causal barriers and not as unidirectional surfaces.

\subsubsection{Penrose diagram}\label{Sec:Causal-structure}

In view of the analysis of the previous sections,  the  Penrose diagram suitably reflecting the causal structure of the Lorentzian-Euclidean Schwarzschild geometry   includes only  the asymptotically flat regions I and III of the Kruskal diagram shown in Fig \ref{Fig-Kruskal}.

Bearing in mind Eq.  \eqref{I-III-relations}, we  define, for  $r>2M$, the coordinates $\bar{U}$ and $\bar{V}$  via the standard  formulas 
\begin{align}
\bar{T}=&\frac{1}{2 } \left(\bar{V}+\bar{U}\right),   
\nonumber \\
\bar{R}=&\frac{1}{2 } \left(\bar{V}-\bar{U}\right),
\end{align}    
with 
\begin{align}
\bar{U} &= \mp \ee^{-\bar{u}/4M}, 
\nonumber \\
\bar{V} &= \pm \ee^{\bar{v}/4M}, 
\end{align}    
where the upper (resp. lower) sign holds in the zone I (resp. III), and  $\bar{u}$ and $\bar{v}$ can be read off from Eq. \eqref{null-coord-u-v-1} with $\varepsilon=1$. Since  $\bar{U}$ and $\bar{V}$ span  the entire real line, they can be compactified by using new variables $\hat{U}$ and $\hat{V}$ according to
\begin{align}
\bar{U}=\tan \hat{U}, \qquad \bar{V}= \tan    \hat{V}, 
\end{align}
with $-\pi/2 < \hat{U} < \pi/2$ and $-\pi/2 < \hat{V} < \pi/2$. It is thus easy to see that  the metric \eqref{Lorentzian-KS-metric}, which, we recall, is valid for $r>2M$,  is conformally equivalent to 
\begin{align}
 \widetilde{\dd s}^2_{\rm L}=-\frac{32 M^3}{r} \ee^{-r/2M} \dd \hat{U} \dd \hat{V} + r^2 \cos^2 \hat{U} \cos^2 \hat{V} \dd \Omega^2.
\end{align}

In the ordinary Schwarzschild solution, $\hat{U}$ and $\hat{V}$ range over the domain $-\pi/2 < \hat{U} + \hat{V} < \pi/2$, which arises from  the  condition $r>0$. Here, however, we have to find the admissible region  of $\hat{U}$ and $\hat{V}$
by taking into account the lower bound $r > 2M$. Firstly, it readily follows from the above definitions that $r= 2M$ leads to the equation  $\tan \hat{U} \tan \hat{V}= 0$, which permits to identify the   future and past event horizons $\mathcal{H}^{\pm}$ as: 
\begin{subequations}
\label{H-plus-minus}
\begin{align}
\mathcal{H}^+:& \;\;   \hat{U}=0 \; \Leftrightarrow \; \hat{T}=\hat{R},
\\
\mathcal{H}^-:& \;\;   \hat{V}=0 \; \Leftrightarrow \; \hat{T}=-\hat{R},
\end{align}    
\end{subequations}
where  $\hat{T}$ and $\hat{R}$ read as 
\begin{align}
\hat{T}=& \hat{V}+\hat{U}, 
\nonumber \\
\hat{R}=& \hat{V}-\hat{U}. 
\end{align}  
The inequality $r>2M$  is then mapped to  $\tan \hat{U} \tan \hat{V}< 0$, which confines the causal structure to the two diamond-shaped domains 
\begin{align}
(\widetilde{{\rm I}})&: -\pi/2 < \hat{U}<0, \, 0<\hat{V}<\pi/2,
\label{tilde-I-bis}
 \\
 (\widetilde{{\rm III}})&: 0 < \hat{U}<\pi/2, \, -\pi/2<\hat{V}<0,
 \label{tilde-III-bis}
\end{align}
or, equivalently, 
\begin{align}
(\widetilde{{\rm I}}):&  -\hat{R}<\hat{T}<\hat{R}, \, \hat{R}-\pi <\hat{T} < - \hat{R}+ \pi, \, 0 < \hat{R}<\pi, 
\label{tilde-I}
\\
(\widetilde{{\rm III}}):&  \hat{R}<\hat{T}<-\hat{R}, \, -\hat{R}-\pi <\hat{T} <  \hat{R}+ \pi, \, -\pi < \hat{R}<0.
\label{tilde-III}
\end{align}
The regions  $\widetilde{{\rm I}}$ and $\widetilde{{\rm III}}$  can be interpreted as the compactified versions of the patches I and III of the Kruskal diagram. They are bounded by  $\mathcal{H}^{\pm}$, and  the  lines $\hat{T}=\mp \hat{R}+ \pi$ (or, equivalently,  $\hat{V}=\pi/2$ and $\hat{U}=\pi/2$) and $\hat{T}=\pm  \hat{R}-\pi$ (or $\hat{U}=-\pi/2$ and $\hat{V}=-\pi/2$), which correspond to  $\mathscr{J}^+$ and $\mathscr{J}^-$, respectively.   
\begin{figure*}[bht!]
\centering\includegraphics[scale=0.30]{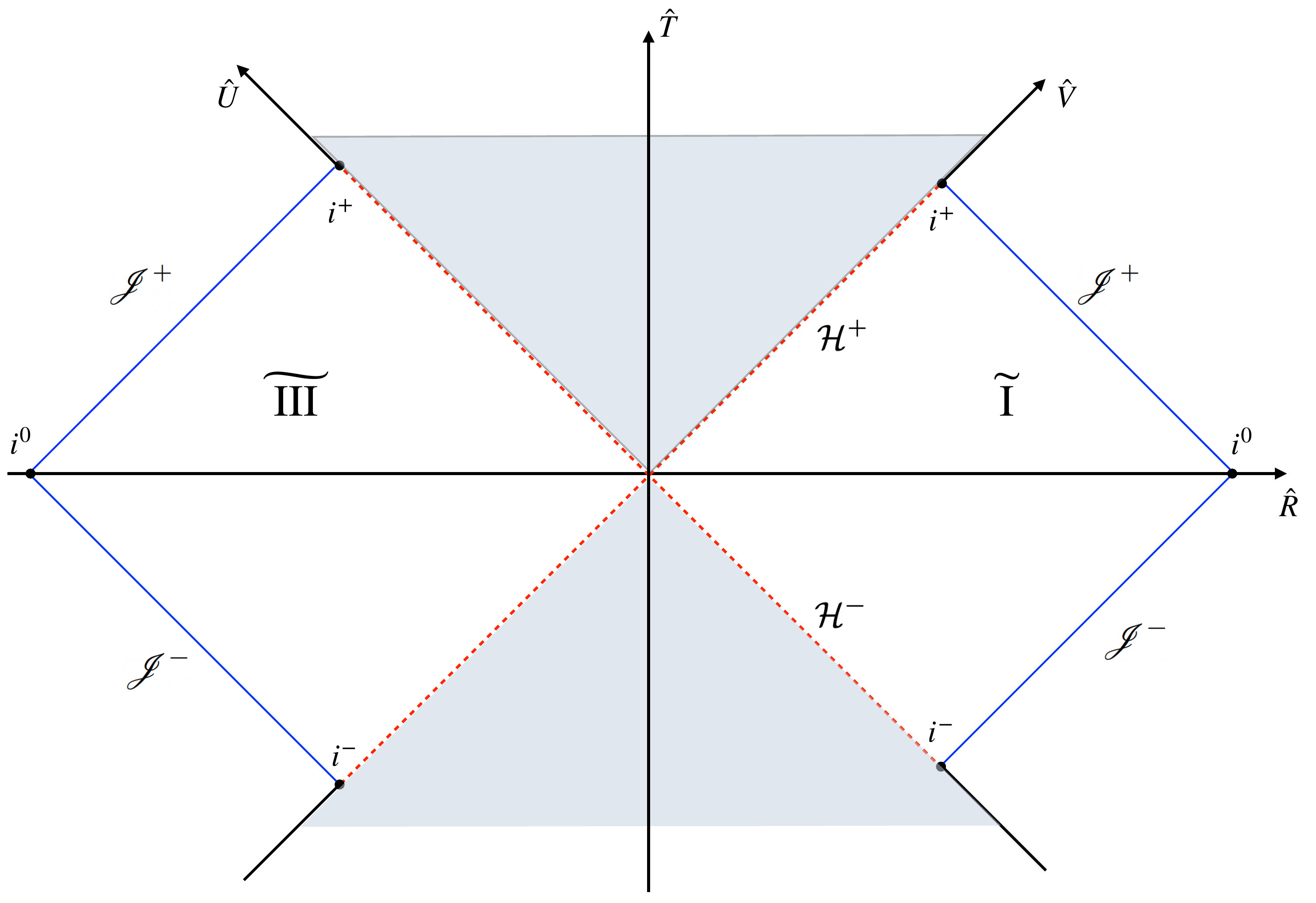}\hspace{1.20cm}
\caption{Penrose diagram of the Lorentzian-Euclidean Schwarzschild geometry. The Lorentzian regions $\widetilde{{\rm I}}$ and $\widetilde{{\rm III}}$ (see Eqs. \eqref{tilde-I-bis}--\eqref{tilde-III}) correspond to the outer domains of the Kruskal diagram of Fig. \ref{Fig-Kruskal}.  The event horizons $\mathcal{H}^{\pm}$ (see Eq. \eqref{H-plus-minus}), represented as red dashed lines, act as causal boundaries rather than unidirectional surfaces. No geodesic can enter or emerge from the shaded Euclidean regions beyond $\mathcal{H}^{\pm}$, which thus lie outside the causal structure of the spacetime and are excluded from the diagram. The resulting spacetime is geodesically complete.}
\label{Fig-Penrose-Schwarzschild}
\end{figure*}

The Penrose diagram is displayed in Fig. \ref{Fig-Penrose-Schwarzschild}. Its structure reflects that the spacetime is asymptotically flat, geodesically complete, and  consists of two causally disconnected regions, with no causal extension beyond the horizons being possible due to the metric degeneracy. Although  appearing as  45-degree lines, $\mathcal{H}^{+}$ and $\mathcal{H}^{-}$ correspond to surfaces where matter and photons stop asymptotically, which  justifies our choice  to depict them as red dashed lines.  As anticipated before, this means that, in our model, $\mathcal{H}^{\pm}$  act as    asymptotic boundaries, rather than  one-way membranes. 

The (future and past) timelike infinities   $i^\pm$ contain the future and past endpoints of  all infinitely extended timelike  geodesics that remain entirely within the Lorentzian domain and do not traverse the  horizons. Their definition differs from that in the conventional Schwarzschild case, where timelike geodesics can cross the event horizon and reach $r=0$, and  ties in with the fact that both  the black and white hole singularities can be avoided. Moreover, $\mathscr{J}^{\pm }$ denote the future and past endpoints of complete outgoing and ingoing radial null geodesics that asymptotically reach $r=2M$. 

Some differences emerge with respect to the flat case discussed in Sec. \ref{Sec:Penrose-flat}. Indeed,  our analysis  has revealed  that $i^{\pm}$, along with  a portion of $\mathscr{J}^{\pm}$, do not  belong to the Penrose diagram in Fig. \ref{Fig-Penrose-flat}, unlike the curved solution of Fig. \ref{Fig-Penrose-Schwarzschild}. The main reason is that the change surface $r=2M$ of the  Lorentzian-Euclidean Schwarzschild geometry is (in the Lorentzian terminology) a null surface and hence appears as an inclined straight line in Fig. \ref{Fig-Penrose-Schwarzschild}, while in the Lorentzian-Euclidean flat framework $r=r_0$ is timelike. In some sense, we can thus say that, in the curved situation, the two-spheres $\mathscr{A}^{+}$ and  $\mathscr{A}^{-}$ defined before boil down  to the points $i^+$ and $i^-$, respectively. Furthermore, the  geometry of the flat scenario becomes  Euclidean rather than ultrahyperbolic for $r<r_0$. In this regard, it is worth mentioning that in the literature conformal diagrams for ultrahyperbolic (also referred to as Kleinian) metrics have been studied in  e.g. Refs. \cite{Atanasov2021,Crawley2021}. However, the approach adopted in those works is different from ours, in which the change of signature affects only the temporal component of the metric   without involving its angular part. 

Nonetheless, a crucial  analogy between the flat and curved Lorentzian-Euclidean  models remains. In fact, in  both patterns,  the metric degeneracy prevents any causal communication between the  Lorentzian and Euclidean regions, due to the lack of real null trajectories in the latter domain. This facet can be seen as a key distinctive feature of degenerate metrics with changing signature.

\subsubsection{Properties of the event horizon}\label{Sec:properties-event-horizon}

As  previously discussed, the event horizon acts as a causal barrier rather than a one-way membrane. In this section, we highlight both similarities and differences with the standard general relativistic pattern.

Recalling that the Lorentzian-Euclidean Schwarzschild geometry is static and asymptotically flat, we note that, exactly like in Einstein theory \cite{Morris1988},  the metric component $g_{tt}$ vanishes at $r=2M$, indicating that the proper time lapse becomes zero over any finite coordinate time interval at the horizon. Consequently, mirroring  the ordinary case \cite{MTW}, $r=2M$ defines a two(-plus-one)-dimensional surface, since 
\begin{align}
    \int \limits_{r=2M} \sqrt{\vert g_{tt}g_{\theta \theta} g_{\phi \phi}\vert } \dd t \dd \theta \dd \phi &=0,
\\ \nonumber
  \int \limits_{r=2M,t=const} \sqrt{\vert g_{\theta \theta} g_{\phi \phi}\vert }  \dd \theta \dd \phi &=4 \pi \left(2M\right)^2,
\end{align}
and, in addition, it is null, because $g^{rr} \left(r=2M\right)=0$. Therefore,  the event horizon  can still be considered as a \qm{point} of no return, although with a slightly different nuance than in Einstein gravity. In fact, in the conventional Schwarzschild solution, once a causal geodesic  \emph{crosses} the horizon, it can no longer escape to infinity, since when $r<2M$ all future-directed paths go in the direction of decreasing $r$ \cite{Carroll2004}. On the other hand, in our model, once a causal geodesic evolving in the Lorentzian domain $\mathscr{D}_+$ (asymptotically) \emph{reaches} $r=2M$, the dynamics cannot proceed beyond it, thereby allowing the singularity to be avoided. Bearing in mind this detail, the event horizon can be defined as usual \cite{Hawking1973-book,Wald-book1984}, i.e.,  as  the boundary (of the closure) of the causal past of $\mathscr{J}^+$ (similar considerations easily apply to the past white hole horizon $\mathcal{H}^-$).  However, a caveat should be mentioned: strictly speaking, the future and past event horizons can no longer be regarded as the boundaries of the black and white hole regions, respectively, as these domains are excluded from the Penrose diagram in Fig. \ref{Fig-Penrose-Schwarzschild}. Despite this, also in the Lorentzian-Euclidean Schwarzschild framework the  zones I and III, or equivalently $\widetilde{{\rm I}}$ and $\widetilde{{\rm III}}$,  remain causally disconnected, although the underlying  reasons  differ from those  in the ordinary Schwarzschild geometry. In addition, patches II and IV continue to be  isolated from the outside and hence invisible to an external observer.

We can thus conclude that many features associated to the event horizon are maintained in the Lorentzian-Euclidean Schwarzschild spacetime,  in spite of the aforementioned distinctions.

\section{The Matter Accretion } \label{Sec:Accretion-Matter}


Accretion of matter onto  black holes is a process of long-standing interest in astrophysics (see e.g. Refs. \cite{Malec1999,Babichev2005,Babichev2013,Martinez2014,Sharif2016,Salahshoor2018,Nozari2020,Akbarieh2023,Mustafa2023c,Fathi2023c,Nozari2024,Murtaza2024,Murtaza2024b,Mustafa2025,Akbarieh2025} and references therein). Rotating gaseous material  in unstable bounded orbits is drawn toward  black holes and leads to  the formation of accretion disks.  This mechanism transfers energy into the surrounding environment, giving rise to spectacular phenomena such as powerful jets, high-energy radiation, and quasars \cite{Martinez2014}. 

The geodesic motion of particles near black holes,  particularly  photon trajectories and innermost stable circular orbits, plays a crucial role in characterizing the properties of accretion disks. Furthermore, these structures  serve as  valuable probes of gravity in the strong-field regime and provide a means to distinguish between different   models of black holes.  

For these reasons, after  outlining the general features of massive and massless particle dynamics, we  now explore the    matter accretion in the  Lorentzian-Euclidean geometry \eqref{Lorentzian-Euclidean-Schwarzschild}. In our setup, the positions of stable and unstable circular orbits remain the same as in the corresponding Lorentzian framework, with the main difference   that our black hole gathers matter around the event horizon, without swallowing it. We investigate how this novel feature  affects the equations governing the accretion dynamics in Sec. \ref{Sec:dynamical-equations-matter}, while   Sec. \ref{Sec:Mass-evolution} addresses the  evolution of the black hole mass.

\subsection{Accretion dynamics}\label{Sec:dynamical-equations-matter}


In this section, we will deal with the accretion of matter  by considering an isotropic perfect fluid  that is supposed to flow at  constant temperature. For such kind of fluid, referred to as isothermal, the pressure and density are proportional, and  the speed of sound  remains constant throughout the accretion. 

The proper definition of the  stress-energy tensor of an ideal fluid should involve  measurements  made by
an experimenter   comoving with the fluid element. For this reason, the standard procedure consists in resorting to the local rest frame of this observer (see Sec. \ref{Sec:Local-Lorentzian-Euclidean-transf}), where  one finds $T^{ab}={\rm diag}(\rho,p,p,p)$, $\rho$ and $p$ being the fluid total energy density and isotropic pressure, respectively \cite{Rezzolla2013}.  If $U^a=(1,0,0,0)$ is the fluid four-velocity normalized to  (cf. Eq. \eqref{flat-metric})
\begin{align}
f_{ab}U^aU^b = - \varepsilon,
\label{fluid-flat-norm}
\end{align}
then,  in the fluid element Lorentz rest frame, we can write
\begin{align}
 T^{ab} =    \left(p/\varepsilon +\rho\right) U^a U^b + p f^{ab},
\label{EMT-flat}
\end{align}
while the covariant components read as $T_{ab}={\rm diag}(\varepsilon^2 \rho,p,p,p)$. Due to the $\varepsilon$-dependent  relation \eqref{fluid-flat-norm}, in our settings $T^{ab}U_a U_b=  \varepsilon^2 \rho$,  which shows that the physical (proper) energy density, locally evaluated by the comoving observer, is $ \rho_{\rm phys} :=\varepsilon^2 \rho$. As a consequence, the rest-mass density current vector assumes the form 
\begin{align}
J^a=\varepsilon^2 \rho U^a,
\label{J-vector}
\end{align} 
a  result that will be crucial in our forthcoming analysis.

By exploiting the principle of  general covariance, it follows from Eq.  \eqref{EMT-flat} that the energy-momentum tensor of a perfect fluid can be expressed, in any reference system, as 
\begin{align}
T^{\mu \nu}= \left(p/\varepsilon +\rho\right) U^\mu U^\nu + p g^{\mu \nu}.
\label{EMT-epsilon}
\end{align}

We will   derive the dynamical equations pertaining to the relativistic spherically symmetric stationary accretion of matter  by employing Eq. \eqref{EMT-epsilon}. We provide a general overview of the process in Sec. \ref{Sec:general-setup}, and then proceed to examine the Lorentzian-Euclidean geometry in Sec. \ref{Sec:Our-setup}.

\subsubsection{The general setup}\label{Sec:general-setup}

In this section, we describe the matter  accretion in a generic static and spherically symmetric  black hole geometry 
\begin{align}
\dd s^2= \mathscr{G}_{\mu \nu} \dd x^\mu \dd x^\nu=-h(r) \dd t^2 + \frac{\dd r^2}{f(r)} + r^2 \dd \Omega^2,
\label{generic-spherical-metric}
\end{align}
where $h(r)$ and $f(r)$ are generic functions of the radial variable $r$. 

We model the accreting material as a perfect fluid with stress energy-tensor given by Eq. \eqref{EMT-epsilon}, and suppose that the temporal component $\mathscr{G}_{tt}$ of  metric \eqref{generic-spherical-metric} depends on a function $\varepsilon(r)$  having similar properties as in Eq. \eqref{epsilon-of-r}. This ensures that  $\mathscr{G}_{\mu \nu}$ becomes degenerate at some point and  exhibits a varying signature. Under these hypotheses, we can thus  assume that the fluid four-velocity $U^\mu$ complies with the pointwise normalization law (cf. Eq. \eqref{velocity-norm})
\begin{align}
\mathscr{G}_{\mu \nu} U^\mu U^\nu = - \varepsilon(r).
 \label{pointwise-relation}
 \end{align}
In this way,  the extension  to the Lorentzian-Euclidean black hole \eqref{Lorentzian-Euclidean-Schwarzschild} will  arise naturally (see Sec. \ref{Sec:Our-setup} below).  

By symmetry arguments, we can restrict our study to the equatorial plane $\theta=\pi/2$. In addition, assuming the fluid is flowing radially, the general form of its four-velocity becomes
\begin{align}
 U^\mu=\left(U^t,U^r,0,0\right),    
\end{align}
where, for forward-in-time flow, $U^t$ must be positive, and for  inward accretion, we have $U^r<0$. Then,  the identity  \eqref{pointwise-relation}  permits to write 
\begin{align}
 U^t = \sqrt{\frac{\varepsilon f + \left(U^r\right)^2}{h f}}.
 \label{accretion-U-t}
\end{align}
Bearing in mind the above formula,  from  the time component of the conservation relation $\nabla_\mu T^{\mu 0}=0$ and the continuity equation $U_\mu \nabla_\nu T^{\mu \nu}=0$, we obtain, after some calculations,
\begin{widetext}
\begin{align}
 & \left( p^\prime  + \varepsilon \rho^\prime\right)  U^r \left[\left(U^r\right)^2 + \varepsilon f\right]  +   \left(p + \varepsilon \rho \right) \Biggl\{U^r  \left( \varepsilon f \frac{h^\prime}{h} +2 \varepsilon \frac{f}{r}   -\frac{1}{2} \varepsilon  f^\prime   \right) + \left( U^r \right)^3  \left(\frac{h^\prime}{h}-\frac{f^\prime}{f}+\frac{2}{r}\right) 
\nonumber \\
& \quad + \left(U^r\right)^\prime  \left[\varepsilon f + 2 \left(U^r\right)^2\right] \Biggr\}+\frac{1}{2} \left(-p + \varepsilon \rho \right)  \varepsilon^\prime f  U^r -p \frac{\varepsilon^\prime}{\varepsilon}\left(U^r \right)^3=0, \label{accretion-1}
\\
 &\frac{\varepsilon \rho^\prime}{p + \varepsilon \rho}+ \frac{1}{2}\left(\frac{-p+\varepsilon \rho}{p + \varepsilon \rho}\right)\frac{\varepsilon^\prime}{\varepsilon} + \frac{1}{2} \left(\frac{h^\prime}{h}-\frac{f^\prime}{f}\right)+ \frac{2}{r}+ \frac{\left(U^r\right)^\prime}{U^r}  =0,
 \label{accretion-2}
\end{align}
\end{widetext}
respectively, where hereafter a prime denotes differentiation with respect to the radial variable. It is worth noticing that, in the limits $h=f=1-2M/r$, $\varepsilon=1$ and $\varepsilon^\prime =0$, one easily recovers, from Eqs. \eqref{accretion-1} and \eqref{accretion-2},  the conventional General Relativity equations  (see e.g. Ref. \cite{Babichev2005}). In the general setup under investigation,  Eqs. \eqref{accretion-1} and \eqref{accretion-2} yield the integrals of motion
\begin{align}
&\left(\frac{p}{\varepsilon} + \rho \right) r^2 U^r  \frac{h}{f} \sqrt{\varepsilon f + \left(U^r \right)^2} = A_0,
\label{accretion-3}
\\
& r^2 U^r \sqrt{\frac{h}{f}} \, {\rm exp} \left[\int \frac{\varepsilon \dd \rho}{p + \varepsilon \rho}+\frac{1}{2}\int \left(\frac{-p+\varepsilon \rho}{p + \varepsilon \rho} \right)\frac{\dd \varepsilon}{\varepsilon}\right] =A_1,  
 \label{accretion-4}
\end{align}
respectively,  $A_0$ and $A_1$ being  integration constants. In the standard Lorentzian case with $\varepsilon=1$, the second integral in Eq. \eqref{accretion-4} vanishes, while the expression occurring in the first one reduces to $\dd \rho /\left(p + \rho \right) \equiv \dd n/n$, with $n$ denoting the particle number density. The joint use of Eqs. \eqref{accretion-3} and \eqref{accretion-4} leads, for a generic function $\varepsilon(r)$,   to
\begin{align}
& \left(\frac{p}{\varepsilon} + \rho \right) \sqrt{h \left[\varepsilon+\frac{\left(U^r\right)^2}{ f}\right]} \;  {\rm exp} \biggl[-\int \frac{\varepsilon \dd \rho}{p + \varepsilon \rho}
\nonumber \\
&-\frac{1}{2}\int \left(\frac{-p+\varepsilon \rho}{p + \varepsilon \rho} \right)\frac{\dd \varepsilon}{\varepsilon}\biggr]= A_2,
\label{accretion-5}
\end{align}
with 
\begin{align}
A_2 := A_0/A_1. 
\label{A2-def}
\end{align}
Owing to  Eq. \eqref{J-vector}, the equation of mass flux is slightly modified and reads as $\nabla_\mu \left(\varepsilon^2 \rho U^\mu \right)=0$, which gives
\begin{align}
  r^2 U^r \varepsilon^2 \rho \sqrt{\frac{h}{f}}= A_3,  
\label{accretion-6}
\end{align}
where $A_3$ is an integration factor that is inherently negative for radially infalling matter (for which $U^r <0$). It then follows, from Eqs. \eqref{accretion-4}--\eqref{accretion-6}, that
\begin{align}
\left(\frac{ p+\varepsilon\rho}{\varepsilon^3 \rho}\right) \sqrt{ h \left[\varepsilon+\frac{\left(U^r\right)^2}{ f}\right]}  = A_4,
\label{accretion-7}
\end{align}
with
\begin{align} \label{A4-A0-A3-relation}
    A_4 := A_0/A_3.
\end{align}

It is important to emphasize that, in keeping with the  model reliance on $\varepsilon$, the constants $A_i$ ($i=0,1,2,3,4$) might, in principle, depend on $\varepsilon$. As shown before, something similar  occurs in massive and massless geodesic dynamics for the integrals of motion $E$ and $\Omega$ (cf. Eqs. \eqref{E-alpha-varepsilon-2} and \eqref{Omega-expression}).

At this stage, an equation of state linking $p$ and $\rho$ is called for. A first hint about its form can be obtained from  the zero-trace condition of the stress-energy tensor \eqref{EMT-epsilon}, which, in the domain where  $\varepsilon \neq 0 $, yields $p=(1/3) \varepsilon \rho$. Since this relation corresponds to the radiation-matter case, we can propose an identity of the form:
\begin{align}
p = \eta \varepsilon \rho,
\label{EoS}
\end{align}
where $\eta$ is the equation-of-state parameter, which reproduces the radiation-field scenario when $\eta=1/3$. However, as it stands, Eq. \eqref{EoS}  cannot be valid in the whole spacetime because it was  derived under the assumption  $\mathscr{G}_{\mu \nu} \mathscr{G}^{\mu \nu} =4$. Due to the degenerate character of the metric, this formula  should instead be replaced by    $\mathscr{G}_{\mu \nu} \mathscr{G}^{\mu \nu}= 4 - m$, where $m$ is the dimension of the kernel of the linear application  $\mathcal{G}: v^\mu \mapsto \mathscr{G}_{\mu \nu} v^\nu$, which maps contravariant vectors to covariant ones (see e.g. Ref. \cite{Gunther} for further details). Thus, Eq. \eqref{EoS}  should be slightly modified to also accommodate cases where $\varepsilon$ vanishes, with  the new ensuing version  reflecting the jump discontinuity in the trace $\mathscr{G}_{\mu \nu} \mathscr{G}^{\mu \nu}$ of the metric. A reasonable choice  is to allow  $\eta$  to depend on $\varepsilon$ via the linear function 
\begin{align}
\eta = \tilde{\eta} \varepsilon,
\label{sigma-relation}
\end{align}
which  leads to $p = \tilde{\eta} \varepsilon^2 \rho = \tilde{\eta}   \rho_{\rm phys}$. This revised formulation has the advantage of establishing a relationship between the physical pressure and energy density, and it is unaffected by the sign of $\varepsilon$.  Consequently, when applied  to the Lorentzian-Euclidean Schwarzschild black hole, it retains the same functional form both inside and outside the event horizon.

Bearing in mind the equation of state \eqref{EoS} supplemented by Eq. \eqref{sigma-relation},  we can derive, from Eq. \eqref{accretion-7}, the following expression for the radial component of the four-velocity of the accreting fluid:
\begin{align}
U^r= - \left(\frac{1}{\eta+1}\right) \sqrt{\frac{\varepsilon f}{h}} \sqrt{\varepsilon^3 \left(A_4\right)^2 -  h \left(\eta +1 \right)^2},  
\label{accretion-8}
\end{align}
which, when plugged into Eq. \eqref{accretion-6}, gives, for the energy density, the following expression
\begin{align}
 \rho= - \frac{A_3}{\varepsilon^2 r^2} \frac{\left(\eta+1\right)}{\sqrt{\varepsilon}\sqrt{\varepsilon^3 \left(A_4\right)^2 -  h \left(\eta +1 \right)^2}},
\label{accretion-9}
\end{align}
where we recall  that in our hypotheses $A_3$  is negative. Finally, the knowledge of $\rho$ permits to obtain $p$ from the equation of state \eqref{EoS}. 

Having obtained a closed form expression for the dynamical parameters $U^r$, $\rho$, and $p$, the problem of describing the accreting fluid motion in the generic geometry \eqref{generic-spherical-metric} is formally solved. 

\subsubsection{The Lorentzian-Euclidean setup}\label{Sec:Our-setup}


The general analysis of the previous section relies on Eq. \eqref{generic-spherical-metric} and hence it applies to any degenerate, signature-changing, spherically symmetric metric. Thus, we can now easily pass to the Lorentzian-Euclidean geometry \eqref{Lorentzian-Euclidean-Schwarzschild}. In fact, in this case,  from Eq. \eqref{accretion-8} we find 
\begin{align}
U^r= -\frac{  \sqrt{\varepsilon}}{\left(\eta+1\right)} \sqrt{ \left(\varepsilon A_4\right)^2 -  \left(1-2M/r \right)  \left(\eta +1 \right)^2 },
\label{radial-fluid-1}
\end{align}
which shows that, differently from the Einstein theory, the radial fluid velocity vanishes on the black hole event horizon and becomes imaginary inside it. This is a significant result, since the same behavior is observed for both  the freely falling and accelerated particles (see Secs. \ref{Sec:Radially-infall-particles} and \ref{Sec:radial-photons}; the  case of radially accelerated  observers has been examined in Ref. \cite{Capozziello2024}). 

It is known that the hydrodynamics equations of perfect fluids establish that the motion becomes geodesic  in the case of  uniform or zero pressure  \cite{Rezzolla2013}. Consequently, comparing Eq.   \eqref{radial-fluid-1} with $\eta=0$, i.e.,  
\begin{align}
    \left. U^r \right \vert_{\eta=0}= -\sqrt{\varepsilon}\sqrt{ \left(\varepsilon A_4\right)^2 -  \left(1-2M/r \right)  },
\end{align}
and the  velocity of an infalling observer  in radial free fall  (see Eq. \eqref{free-fall-velocity}), leads, upon recalling the relation  \eqref{E-alpha-varepsilon-2} for the particle energy $E$, to the conclusion that $A_4$ implicitly involves a factor $\varepsilon^{-1}$. We can thus write
\begin{align}
A_4 = \frac{\tilde{A}_4}{\varepsilon},
\label{A4-new}
\end{align}
which in turn implies 
\begin{align}
A_3 = \tilde{A}_3 \varepsilon,
\label{A3-new}
\end{align}
where this last identity has been derived from Eq. \eqref{A4-A0-A3-relation} by  assuming that $A_0$ does not depend on $\varepsilon$. Notice that Eq. \eqref{A3-new} guarantees that the (negative) sign of  $A_3$  is not altered for $r>2M$, i.e., the region of the spacetime where the accreting fluid dynamics is allowed.

Bearing in mind Eqs. \eqref{accretion-9} and \eqref{EoS},  we find that, in the Lorentzian-Euclidean Schwarzschild model, the physical density and pressure take the form
\begin{align}
 \rho_{\rm phys} & = - \frac{A_3}{r^2} \frac{\left(\eta+1\right)}{\sqrt{\left(\varepsilon^2 A_4\right)^2 - \varepsilon^2 \left(1-2M/r \right)  \left(\eta +1 \right)^2 }},
 \label{rho-phys-final}
\\
p &= - \frac{A_3}{r^2} \frac{1}{\varepsilon} \frac{\eta\left(\eta+1\right)}{\sqrt{\left(\varepsilon^2 A_4\right)^2 - \varepsilon^2 \left(1-2M/r \right)  \left(\eta +1 \right)^2 }},
\label{pressure-final}
\end{align}
respectively. By virtue of Eqs. \eqref{A4-new}  and \eqref{A3-new}, $\rho_{\rm phys} :=\varepsilon^2 \rho$  remains finite on $r=2M$,  attains real values for $r<2M$,  and diverges at $r=0$, as in Einstein theory. The pressure also exhibits the same behavior,  provided  we further take into account Eq.   \eqref{sigma-relation}. The fact that $\rho_{\rm phys}$ and $p$  do not blow up at $r=2M$   ties in with the key peculiarities of the Lorentzian-Euclidean black hole. This object in fact accumulates matter, which, as pointed out before,  takes an infinite time to reach the event horizon.  On the other hand, had the matter been able to get to the event horizon within a finite amount of time, the density and pressure would have been expected to diverge  at $r=2M$.

\subsection{Mass evolution} \label{Sec:Mass-evolution}

The rate of change of the black hole mass due to the accretion process can be obtained by integrating the flux of the fluid over the surface of the black hole via the relation \cite{Salahshoor2018}
\begin{align}
    \frac{\dd M}{\dd t}= \int  \sqrt{-g}\, \dd \theta \dd \phi T^r_t.
\end{align}
Upon exploiting Eqs. \eqref{EMT-epsilon}, \eqref{generic-spherical-metric},  and \eqref{accretion-U-t}, we arrive at 
\begin{align}
\frac{\dd M}{\dd t}=- 4 \pi r^2  \left(\frac{p}{\varepsilon} + \rho\right) \frac{h}{f} U^r \sqrt{\varepsilon f + \left(U^r \right)^2} = - 4 \pi A_0,  
\label{mass-rate}
\end{align}
where, in the last equality, we  used Eqs. \eqref{accretion-4}--\eqref{A2-def}. Upon plugging  Eqs. \eqref{radial-fluid-1},  \eqref{rho-phys-final}, and \eqref{pressure-final} into the above formula, we find that $\dd M / \dd t $ does not depend on $\varepsilon$, in agreement with our hypothesis that $A_0$ does not involve an implicit $\varepsilon$ factor. It follows, from Eq. \eqref{mass-rate}, that the time derivative of $M$ is positive when $U^r <0$, thereby showing that the mass of the black hole increases as it accretes ordinary matter, like in the standard Lorentzian model.

Since the Lorentzian-Euclidean geometry \eqref{Lorentzian-Euclidean-Schwarzschild} is identical, outside the event horizon, to the conventional Schwarzschild solution, the integration process of  Eq. \eqref{mass-rate} can be performed following the standard arguments outlined in Refs. \cite{Babichev2005,Babichev2013}. Therefore, under the assumption of steady-state, far-field accretion and  by writing $A_2= (p_\infty + \rho_\infty) \sqrt{h_\infty}$,  Eq. \eqref{mass-rate} gives 
\begin{align}
\frac{\dd M}{\dd t}=- 4 \pi A_1   (p_\infty + \rho_\infty) \sqrt{h_\infty} M^2,
\label{mass-rate-2}
\end{align}
 the asymptotic quantities $p_\infty$,  $\rho_\infty$, and  $h_\infty $ being evaluated at large distance from the black hole (i.e., for $r \to \infty$).  Owing to the isothermal fluid assumption, $p_\infty$ and $ \rho_\infty$  can be  treated as constants. Therefore, from Eq. \eqref{mass-rate-2}, we easily obtain the instantaneous black hole mass as a function of the initial value $M_i$: 
\begin{align}
M(t)= \frac{M_i}{1-t/\mathcal{T}},    
\end{align}
with the critical time $\mathcal{T}$ marking the instant when $M(t)$ blows up and being given by 
\begin{align}
\mathcal{T}= -\left[4 \pi M_i A_1 (p_\infty + \rho_\infty) \sqrt{h_\infty} \right]^{-1},
\end{align}
where we recall  that $A_1$ is intrinsically negative for infalling matter (cf. Eq. \eqref{accretion-4}). 

The above calculations are identical to those derived in the Einstein theory \cite{Babichev2005} and hence prove that the Lorentzian-Euclidean black hole is able to accumulate mass in its exterior region, like in the standard Lorentzian world. This is a remarkable result, especially in light of the dynamical behavior of the perfect fluid, whose motion halts at the event horizon, as discussed in the previous section. 

An important remark is necessary at this point. A  further scenario to be taken into account  is the radial accretion in the framework of the   Michel model, which  describes a  steady-state, spherically symmetric,  transonic inflow of matter onto a Schwarzschild black hole  \cite{Michel1972} (for  recent applications see e.g. Refs. \cite{Chaverra2015,Abbas2020,Aguayo-Ortiz2021}). In the standard situation, the infalling fluid accelerates inward and crosses the event horizon, continuing towards the central singularity. In contrast, in the Lorentzian-Euclidean geometry, the radial  flow  is  halted at the horizon, with its velocity asymptotically approaching zero. The infalling matter thus accumulates at $r=2M$ without penetrating further. In conclusion, the Michel model  is thus  ideal to probe  the modified  causal structure discussed in Sec. \ref{Sec:Penrose-diagrams}.

\section{Discussion and conclusions  }
\label{Sec:Conclusion}

Black hole singularities represent a drawback of General Relativity: although they are hidden behind horizons and thus inaccessible to direct observations, such as  gravitational-wave detections or Event Horizon Telescope imaging, they nonetheless mark the boundary where the theory ceases to be predictive and physically well-defined. From a theoretical perspective, any solution containing singularities motivates the search for alternative, geodesically complete spacetimes. The  Lorentzian-Euclidean black hole aims to address precisely this issue. Motivated by the fact that any assumption about physics or dynamics  inside the event horizon  remains purely conjectural,  and that definite statements can only be made in the observable exterior region, our model  preserves the  features of the standard  Schwarzschild spacetime in the domain $r>2M$, with  modifications occurring in the interior unobservable zone $r<2M$ and at near-horizon scales.

The Lorentzian-Euclidean black hole is an event-horizon degenerate, signature-changing solution of vacuum Einstein equations  where the central singularity can be  dynamically avoided through atemporality.  This mechanism permits the time variable to shift from being real-valued to imaginary and can be related to the boundedness along radial trajectories of the Kretschmann invariant $\mathcal{K}$, which we propose to interpret as the  parameter \qm{measuring} the degree of atemporality. 

Following our previous work \cite{Capozziello2024}, in this paper, we have further delved into the Lorentzian-Euclidean geometry. In Sec. \ref{Sec:generalized-transformation}, we  developed a set of Lorentzian-Euclidean coordinate transformations that fit for  the main properties of the  metric and generalize the ordinary local Lorentz transformations of Einstein theory. This has allowed us to  set up a boost along the radial direction that connects the reference frames of the static and free-fall observers. This examination   represented the starting point for the  subsequent analysis in Secs. \ref{Sec:null-geodesics}, \ref{Sec:Penrose-diagrams}, and \ref{Sec:Accretion-Matter}, where we have addressed the   photons dynamics,  the causal structure of the spacetime, and the matter accretion phenomenon. 

As with massive objects, light rays cannot enter the black hole, as their radial velocity vanishes asymptotically on the event horizon and becomes imaginary  inside it. This proves that neither timelike nor null geodesics  cross the event horizon,   further confirming that the $r=0$ singularity is averted in our model. Therefore,  the concept of atemporality can be extended to causal geodesics, as it introduces a natural cutoff for the radial paths $r(\vartheta)$, which are constrained to satisfy the relation $2M \leq  r(\vartheta) < \infty$, with $\vartheta \equiv \sigma$ for massive particles and  $\vartheta \equiv \lambda$ for massless ones. We highlight that  our approach does not remove the singularity,  but rather  explores how   its dynamical manifestation  can be  avoided within a consistent theoretical  framework.

The analysis of the causal structure  reveals that the event horizon acts as  a causal barrier rather than a one-way membrane,  preventing any causal communication between the  Lorentzian and Euclidean regions. This property can be seen as a fundamental trait of varying-signature degenerate metrics. Strictly speaking, the Penrose diagram consists only of regions I and III of  the Kruskal extension, with regions II and IV  excluded, since causal geodesics cannot cross the degenerate surface $r=2M$ (see Figs. \ref{Fig-Kruskal}-\ref{Fig-Penrose-Schwarzschild}). Nevertheless, many aspects of  the standard definition of the event horizon are preserved.   

The matter accretion onto the Lorentzian-Euclidean black hole mirrors the behavior of freely falling bodies, since the motion of the perfect fluid halts at  $r=2M$. Despite that,  the accretion process and mass evolution proceed like in Einstein gravity, with  the physical density and pressure exhibiting  finite values on the event horizon. 

One of the key strengths of the Lorentzian-Euclidean model is its ability to reproduce established results using  methods distinct from those of conventional frameworks. In fact, although it differs from its Lorentzian counterpart in the near-horizon domain, we have obtained results analogous to those in General Relativity, such as: (i) the energies  $E^{\rm (s)}$ and $ \tilde E^{({\rm s})}$ of a body in radial free fall, as  locally determined by the static observer, blow up at $r=2M$; (ii) the ordinary blueshift formula presented in Eq. \eqref{omega-s}; (iii) the behavior of the frequency $\omega^{(\rm s)}$ of  radially moving photons, as measured by the stationary observer, and $\tilde{\omega}^{\rm (f)}$, as evaluated by the geodetic observer, with the former diverging at the event horizon and the latter remaining finite; (iv) the ability of the black hole to accumulate matter, which can lead to the formation of accretion disks in the outer region.  

Our findings can be framed within the broad context of modern black hole physics, where recent surveys  have furnished unprecedented access to the  horizon-scale phenomena. In particular, the study in Ref. \cite{Ru-Sen2023} presents new high-resolution radio observations of M87$^*$, conducted using global very-long-baseline interferometry at a wavelength of 3.5 mm. This research reveals a ring-like accretion structure surrounding the black hole with a diameter of $8.4 \pm 1.1$ Schwarzschild radii, which is approximately $50\%$ larger than the 1.3 mm ring ($5.5 \pm  0.7$ Schwarzschild radii) detected by the Event Horizon Telescope collaboration \cite{EventHorizonTelescope2019b}. This discrepancy  suggests that millimeter-wavelength emission extends beyond the photon ring, and places new constraints on the mechanisms regulating the matter accretion flow.  Similarly, SgA$^*$ has been extensively monitored. Indeed, results in Ref. \cite{Vagnozzi2022} demonstrate that information provided by the Event Horizon Telescope collaboration supports the Einstein theory predictions in the strong-field regime, though some competing paradigms remain plausible. Therefore, the exploration of black hole models alternative to General Relativity remains a hot topic,  especially in light of the wealth of (present and foreseeable) data.

The condition   $r(\vartheta)  \geq 2M$ characterizing radial geodesics can be thought of as a sort of gravitational uncertainty principle, which prevents probing length scales below the event horizon, just as  the Heisenberg uncertainty principle sets fundamental limits on  physical observations in quantum systems. Whether this hypothesis has a quantum origin, and how  quantum effects may affect the Lorentzian-Euclidean geometry and particles dynamics,  represent significant topics that deserve to be discussed in further researches. 

Another  key aspect of our model is  the presence of  imaginary time, which by its very nature cannot correspond to a measurable entity. If we assume that all physical observables must be described by real-valued quantities, then  time is not  operationally definable beyond the horizon. This marks  the onset of atemporality, where standard notions of measurement and dynamics break down. An instructive analogy comes again from the Heisenberg uncertainty principle, as we argue that for $r<2M$, measurements lose operational significance, and all  meaningful observations must be confined to  the Lorentzian region $r>2M$. Similarly, as we have demonstrated in this paper,  the emergence of the imaginary temporal coordinate signals the end of causal evolution, with the interior black hole region becoming effectively disconnected from the observable exterior.

Before concluding, we put forward a few final remarks. First, we acknowledge that for any new solution to Einstein equations to be considered physically relevant, questions such as stability and formation processes must be addressed. In our case, since the exterior geometry of the Lorentzian-Euclidean black hole coincides with the Schwarzschild solution, both stability properties \cite{Chandrasekhar1985} and formation mechanisms \cite{Hawking1973-book,Shapiro1983} are expected to apply in an analogous way in the region  $r>2M$.  Moreover, in a forthcoming paper, our model  will be extended to Kerr-like black holes, in view of representing more realistic astrophysical objects.

\section*{Acknowledgements}
SC  acknowledges the support of  Istituto Nazionale di Fisica Nucleare (INFN) Sez. di Napoli ({\it Iniziative Specifiche} QGSKY and MOONLIGHT2) and the Gruppo Nazionale di Fisica Matematica (GNFM)  of Istituto Nazionale di Alta Matematica (INDAM).  EB acknowledges the support of  Istituto Nazionale di Fisica Nucleare (INFN), {\it Iniziative Specifiche}  MOONLIGHT2.   SDB acknowledges the project (2021-0567-COSMOS) funded by Cariplo Foundation.
This paper is based upon work from COST Action CA21136 {\it Addressing observational tensions in cosmology with systematics  and fundamental physics} (CosmoVerse) supported by COST (European Cooperation in Science and Technology).

\bibliography{references}

\end{document}